\documentclass[sn-mathphys,Numbered]{sn-jnl}

\usepackage{graphicx}%
\usepackage{multirow}%
\usepackage{amsmath,amssymb,amsfonts}%
\usepackage{amsthm}%
\usepackage{mathrsfs}%
\usepackage[title]{appendix}%
\usepackage{xcolor}%
\usepackage{textcomp}%
\usepackage{manyfoot}%
\usepackage{booktabs}%
\usepackage{algorithm}%
\usepackage{algorithmicx}%
\usepackage{algpseudocode}%
\usepackage{listings}%
\usepackage{gensymb}
\usepackage[normalem]{ulem}

\begin{document}

\title[Article Title]{Analysis of transient and intermittent flows using a multidimensional empirical mode decomposition}


\author{\fnm{Lucas} \sur{F. de Souza}} \email{l251480@dac.unicamp.br}

\author{\fnm{Renato} \sur{F. Miotto}} \email{miotto@fem.unicamp.br}

\author*{\fnm{William} \sur{R. Wolf}} \email{wolf@fem.unicamp.br}

\affil{\orgdiv{School of Mechanical Engineering}, \orgname{University of Campinas}, \city{Campinas}, \postcode{13083-860}, \state{São Paulo}, \country{{Brazil}}}

\abstract{Modal decomposition techniques are important tools for the analysis of unsteady flows and, in order to provide meaningful insights with respect to coherent structures and their characteristic frequencies, the modes must possess a robust spatial support. In this context, although widely used, methods based on singular value decomposition (SVD) may produce modes that are difficult to interpret when applied to problems dominated by intermittent and transient events. Fortunately, specific modal decomposition techniques have been recently developed to analyze such problems. However, a proper comparison between existing methods is still lacking from the literature. Therefore, this work compares two recent methods: the fast adaptive multivariate empirical mode decomposition (FA-MVEMD) and the multi-resolution dynamic mode decomposition (mrDMD). These techniques are employed here for the study of flow databases involving transient and intermittent dynamics. Specifically, the investigated problems include an SD7003 airfoil subjected to deep dynamic stall conditions, and a steady NACA0012 airfoil operating at a transitional Reynolds number. In the former case, the methods are employed to investigate the onset and evolution of the dynamic stall vortex (DSV), while in the latter case, intermittent vortex pairing is analyzed. We show that the combination of a multidimensional EMD with the Hilbert transform provides modes with superior spatial support when compared to the mrDMD, also allowing the characterization of instantaneous frequencies of coherent structures. Moreover, the EMD also condenses a larger amount of information within a single intrinsic mode function (IMF).}




\maketitle

\section{Introduction}\label{sec1}

Extracting relevant information about the dynamics of complex unsteady flows can be challenging since they are often dominated by nonlinear phenomena which may be transient and intermittent. In this context, modal decomposition techniques are presented as an important tool to identify coherent structures in unsteady flows, as well as provide their characteristic frequencies. With such information, it is possible to better understand the flow physics and develop more effective control strategies. However, to serve this purpose, it is necessary that the modes obtained by these techniques present a  
meaningful spatial support so that they can be
interpreted with physical rigor. In this sense, assumptions of stationarity and linearity can 
hinder the application of 
some decomposition techniques to inherently transient and intermittent flows.

Several flow modal decomposition techniques are available in the literature for extracting temporal and spatial features of unsteady flows \cite{Taira_modal_analysis_review}.
Some approaches have a rich mathematical foundation such as global stability \cite{THEOFILIS2003249, Vassilios/annurev-fluid-122109-160705, juniper_theofilis_2014} and resolvent analyses \cite{McKeon_PhysRevFluids, Taira_modal_analysis_review, hamada_wolf_pitz_alves_2023}, which are based on the approximation of the Navier-Stokes equations by a discrete linear operator related to an equilibrium state that satisfies the flow governing equations. When such base state is known, the onset of instabilities can be modeled as a linear phenomenon and information can be extracted with respect to the 
spatial support of the unstable modes and the frequencies related to early stages of the transition process.
However, these methods rely on the definition of a proper stationary base flow, hampering their applications in transient problems.

More recently, data-driven methods have been developed with a deep foundation in linear algebra to provide analyses of results extracted directly from numerical simulations or experimental measurements. In this sense, there is a widespread use of techniques such as proper orthogonal decomposition (POD) \cite{Ribeiro_2017, nikolaidis_2023} and dynamic mode decomposition (DMD) \cite{miotto2022analysis, DMD_vinuesa}, besides their variants \citep{SPOD_sieber, SPOD_towne, HODMD_2017, MrDMD_2016}, which are based on the eigendecompostion of a matrix constructed from the flow properties. These data-driven methods have become popular for studies of unsteady flows due to their capabilities to isolate coherent structures and provide insights about their physical features. 
From its mathematical definition, POD is well-suited for the analysis of statistically stationary flows. The method is based on the principal component analysis (PCA), a statistical procedure that employs orthogonal transformation to convert a set of observations of correlated
variables into a set of linearly uncorrelated spatial modes. The individual modes are ranked by energetic content and, by construction, have minimum variance. Despite its success in the analysis of fluid flows, the method has issues to represent the transport of low-rank traveling structures as it is based on the singular value decomposition (SVD) of a correlation matrix computed from the data.
Consequently, these SVD-based approaches may exhibit limitations when
dealing with translational and rotational invariances of low-rank objects embedded in the data.

The DMD, in turn, consists of finding an optimal linear operator connecting the time-dependent variables, which are usually represented by a collection of snapshots, portraying an approximation to the Koopman operator \cite{Tu_etal_2014}. This is achieved by performing an SVD on the data matrix to produce a low-rank representation which allows for a computationally viable eigendecomposition. In summary, the flow data is transformed into an eigenvalue problem in which the eigenvectors represent the spatial modes and the eigenvalues capture the dynamics of the individual modes in terms of frequency and growth rate \cite{schmid_2010, kutz_et_al}. 
However, being based on SVD, the method also presents limitations with respect to the transport of low-rank objects which can lead to an artificial inflation in the dimensionality of the problem and the need for 
a high-rank representation \cite{kutz_et_al}. Furthermore, the method fails to represent transient behavior, regardless of the truncation rank. In order to overcome its drawbacks and improve the robustness of the method, different variants of the standard DMD algorithm have been developed, such as the high-order DMD \cite{HODMD_2017}, which employs a sliding window to compute the decomposition making the method more robust to noisy data, including the possibility of on-the-fly applications \cite{HODMD_on_the_fly_Schlatter_Vinuesa_2023}, and the multi-resolution DMD \cite{MrDMD_2016} (mrDMD) which is more suitable to characterize intermittent and local phenomena \cite{miotto2022analysis}.

In this context, the application of POD and DMD in problems of a highly transient nature, or dominated by intermittent events with the convection of coherent structures, can lead to modes that are difficult to physically interpret. Thus, these flows can be used as a benchmark for evaluating the performance of data-driven modal decomposition techniques in obtaining relevant flow physics. In this regard,
\citet{Diss_Brener} showed that POD has problems characterizing the transport of structures in dynamic stall. For such cases, the lack of a well-defined mean flow impairs the interpretation of the POD modes, since they essentially represent the oscillatory content of the data around a mean. Also in the context of dynamic stall,
\citet{miotto2022analysis} showed that the conventional DMD produces modes with a scattered spatial support when compared to its multi-resolution variant. In the former, the modes must exist throughout the whole periodic cycle, leading to the creation of spurious modes that serve as cancelling artifacts with destructive interference, as in a Fourier transform of an intermittent signal. On the other hand, the mrDMD modes are more prone to identify, both in space and time, the presence of coherent structures related to the onset and evolution of the dynamic stall vortex. 

The empirical mode decomposition (EMD) emerged as an alternative to better characterize transient and intermittent phenomena in fluid flows. 
The technique is capable of handling nonlinear and non-stationary signals, decomposing the data in intrinsic mode functions (IMFs) which incorporate the oscillatory content of the signal and a residue that represents the main data trend \cite{Huang_1998}. Recently, \citet{Ansell_2020_evolution} used EMD together with linear stability analysis 
to show that the emergence of vortices in the leading edge of the airfoil under a periodic pitching motion 
was linked to the roll-up of Kelvin-Helmholtz instabilities in the shear layer. \citet{Ansell_Karen_2020} also employed the EMD to investigate the evolution of the flowfield during the motion of an 
airfoil under deep stall conditions. They were able to identify Kelvin-Helmholtz instabilities during the initial stages of the movement, as well as the characteristic timescale of the dynamic stall vortex shedding process.

EMD was first developed as a signal processing tool to handle one dimensional data using cubic spline interpolation to create envelopes in the sifting algorithm responsible for extracting the IMFs. The first attempts to extend this method to multiple dimensions consisted in performing 1D EMD on slices of the data. However, this approach suffered from the drawbacks of mode mixing which can introduce discontinuities in the multidimensional IMFs \cite{Wu_2009}. This issue leads to results which cannot be physically interpretable. Wu et al. \cite{Wu_2009} proposed a noise aided approach to solve this problem by performing an ensemble averaging of multiple decompositions of the original data disturbed by a white Gaussian noise. However, the cost of performing a spline-based envelope determination for several ensemble realizations can be prohibitive for large datasets. Different approaches to replace the spline interpolation in the sifting algorithm, and to extend the method for multiple dimensions, have been developed such as radial basis functions for surface interpolation in the 2D EMD \cite{Nunes_2003}, order statistics filters in fast and adaptive 2D EMD \cite{Bhuiyan_etal_2008}, 3D EMD \cite{He_etal_2017}, and the fast and adaptive multivariate EMD (FA-MVEMD) \cite{FA_MVEMD_2019}. Besides the multidimensional extensions, efforts have also been made to handle the decomposition of multivariate data such as the velocity field, which needs more than one component to be fully characterized. In this sense, projection methods are used to carry out the identification of the envelopes among the multiple variables in order to guarantee mode alignment among them \cite{Rehman_Mandic_2010, Hemakom_Mandic_2016, Altaf_2007, Rilling_2008}. More recently, \citet{FA_MVEMD_2019} combined multivariate and multidimensional features to EMD in order to analyze planar and volumetric flowfield data. 

In the present work, we assess the potential of the FA-MVEMD and the mrDMD in producing physically interpretable modes when applied to unsteady flows characterized by transient and intermittent behavior. For this, both methods are applied using two datasets obtained from wall-resolved large eddy simulations (LES). The first case consists of a periodic plunging SD7003 airfoil under deep dynamic stall conditions \cite{miotto2022analysis}, while the second case consists of the flow past a NACA0012 airfoil with a fixed angle of attack, under a transitional Reynolds number \cite{ricciardi_wolf_taira_2022}. The reason for selecting these two problems lies in showing how each method performs under different phenomenological circumstances. For example, in the first case, different structures appear due to the airfoil motion, such as Kelvin-Helmholtz instabilities which are responsible for the onset of dynamic stall, besides the dynamic stall vortex and the trailing edge vortex.
On the other hand, for the stationary NACA0012, Kelvin-Helmholtz instabilities appear downstream of a separation bubble that forms on the airfoil suction side. The flapping motion of this bubble causes intermittent vortex pairing, which in turn affects the transition along the boundary layer and generates different levels of trailing-edge noise.



\section{Methodology}\label{sec2}

\subsection{Large eddy simulations}

High-fidelity simulations are performed on a periodic plunging SD7003 airfoil, and for the transitional flow over a NACA0012 airfoil. O-type grids are employed for both cases and, therefore, the equations are solved in a general curvilinear coordinate system. In order to resolve the most energetic flow scales, wall-resolved LES are conducted. The flow equations are solved using the staggered grid approach presented by Nagarajan {\em et al.} \cite{Nagarajan2003}. Therefore, the numerical methodology employed in the spatial discretization combines the application of sixth-order accurate compact schemes for calculation of derivatives and interpolations on the staggered grids. 

The time integration of the flow equations is performed using an explicit third-order compact-storage Runge-Kutta scheme in regions away from solid boundaries. Near the airfoil surface, an implicit second-order Beam-Warming scheme \cite{beam:78} is applied to overcome the stiffness problem typical of boundary layer grids. Sponge layers and characteristic boundary conditions based on Riemann invariants are applied in the farfield, and adiabatic no-slip boundary conditions are used at the airfoil surfaces. The present numerical tool has been validated against experimental results and high-fidelity numerical simulations of turbulent flows \cite{Nagarajan2003, Bhaskaran, wolf2012}, besides the study of dynamic stall \cite{ramos2019active, miotto2022analysis, miotto2021pitch}.

\subsection{Multidimensional empirical mode decomposition}

In order to separate the different spatiotemporal scales of the events in the investigated flows, we perform the fast and adaptive multivariate empirical mode decomposition (FA-MVEMD) into spanwise-averaged collections of pressure snapshots. The EMD is an adaptive data-driven method for flow analysis ideal for unsteady problems. This method produces different intrinsic mode functions (IMFs) that represent the oscillatory content of a time-dependent field around a zero mean, plus a residue which contains the main trends of the data. Here, we analyze the unsteady pressure field since it allows the characterization of vortical structures and, in a multidimensional case, this process is represented by:
\begin{equation}
\label{eq : EMD}
    p(x,y,t) = \sum_{k=1}^{N} \mbox{IMF}_{k}(x, y, t) + \mbox{Residue}(x,y,t) \mbox{ ,} 
\end{equation}
where $p(x,y,t)$ represents the instantaneous pressure field, $k$ is the IMF index and $N$ is the total number of IMFs.

Among the several available options of EMD algorithms, the FA-MVEMD is selected in this work as it provides multidimensional and multivariate capabilities. The approach has been validated in previous experimental investigations of dynamic stall \cite{Ansell_2020_evolution, Ansell_Karen_2020}. We employ the bivariate version of the algorithm available at \url{https://www.mathworks.com/matlabcentral/fileexchange/71270-fast-and-adaptive-multivariate-and-multidimensional-emd}. The collection of pressure snapshots is used in the first channel while white noise is employed in the second channel. This noise-aided method proved to be efficient in enhancing the decomposition of multivariate EMD algorithms, since the added noise homogeneously fills the frequency domain, taking advantage of the mode alignment property between the channels to yield a better separation of scales among the IMFs \cite{Rehman_mandic_2011, Rehman_2013}. The white noise data is generated with 10\% of the original data power considering time-dependent series in each spatial location.

Owing to the fact that the EMD essentially functions as a quasi-dyadic filter bank, the IMFs obtained are almost band-limited \cite{Rehman_mandic_2011}. Thus, when applying a multidimensional EMD, the low-order IMFs (lower values of $k$) incorporate the high-frequency, small-scale events, while the low-frequency, large-scale phenomena are represented by the higher-order IMFs (higher values of $k$). To quantify the frequencies within the IMFs, a spectral analysis is employed using the Hilbert transform at each spatial location of the IMFs. By doing so, a local analytical signal $Z_{k}(x, y, t)$ can be created by combining the original real-valued signal with its complex value computed from the Hilbert transform, as shown below:
\begin{equation}
    Z_{k}(x, y, t) = \mbox{IMF}_{k}(x, y, t) + i H\left\{ \mbox{IMF}_{k}(x, y, t) \right \} = A_{k}(x, y, t) e^{i \phi_{k} (x, y, t)} \mbox{ .}
\end{equation}
In the equations above, the operator $H$ represents the Hilbert transform of its argument and $i$ is the imaginary number. 
The signal phase $\phi$ and its amplitude \mbox{A} are given respectively by:
\begin{equation}
    \phi_{k} (x, y, t) =  \arctan \left( \frac{H\{ \mbox{IMF}_{k}(x, y, t) \} }{\mbox{IMF}_{k}(x, y, t)}  \right) \mbox{ ,}
\end{equation}
and
\begin{equation}
A_{k} (x, y, t) = \sqrt{\mbox{IMF}^{2}_{k}(x, y, t) + (H\{ \mbox{IMF}_{k}(x, y, t) \})^{2}} \mbox{ .}
\end{equation}
Then, the instantaneous frequency is obtained by taking the time derivative of the signal phase as:
\begin{equation}
    f_{k}(x,y,t) =\frac{\partial \phi_{k} (x, y, t)}{\partial t}  \mbox{ .}
\end{equation}

\begin{figure}[hb!]
\centering
\includegraphics[width=.99\textwidth]{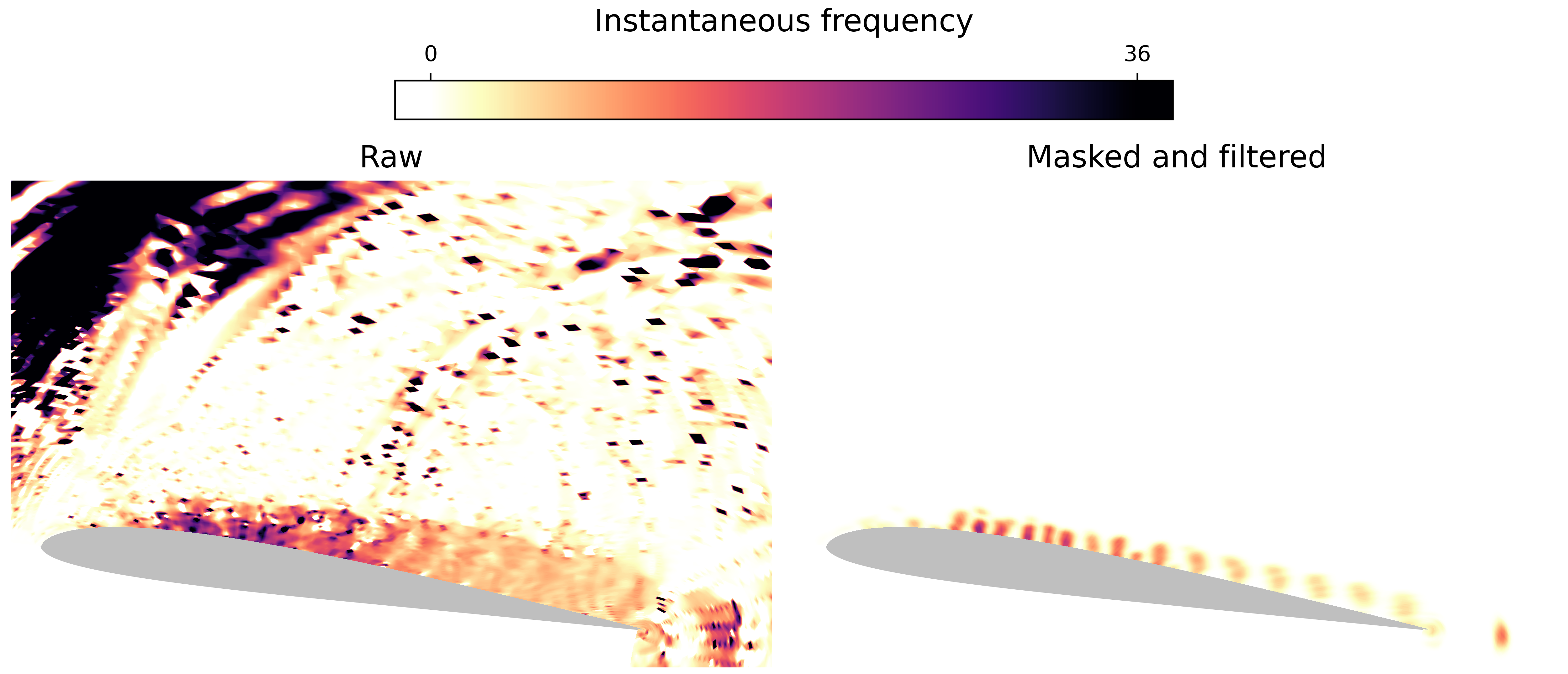}\hfill
\caption{The raw instantaneous frequencies (left) and the post-processed values (right). In the post-processing, amplitudes less than 1\% and positive values are removed from the raw data by the mask-filter, and then a Gaussian filter is applied to smooth out discontinuous interfaces.}
\label{fig : Mask}
\end{figure}

To have more interpretable hydrodynamic features, results from the Hilbert spectral analysis are post-processed with a mask filter, removing the frequencies associated with the stochastic signals and acoustic waves in the near field. For this, we take the absolute value of each IMF and normalize the data by its maximum value. Next, the mask is created to remove instantaneous frequency data where the normalized value of the IMF is within the lowest 1\% or where positive values are found. The reason for removing the data in flow regions where the IMF values are positive is due to the fact that we are interested in studying the dynamics of vortices which are inherently represented by negative pressure fluctuations at their core. A low intensity Gaussian filter is also employed to the contours of the Hilbert spectral analysis to remove eventual rough edges and discontinuities created by the mask filter. It is worth mentioning that this further post-processing does not compromise the data in any form that may lead to a misguided analysis since it only emphasizes the most meaningful features with respect to the amplitude of hydrodynamic events as show in Fig. \ref{fig : Mask}.

\subsection{Multi-resolution dynamic mode decomposition}

Although the present work is focused on the application of the multi-resolution version of the DMD algorithm, the fundamentals of the standard DMD approach are also discussed here as it is used in the mrDMD and can provide insights for its downsides. The DMD~\citep{schmid_2010, Tu_etal_2014} is a technique that isolates the most dynamically significant modes containing low-rank flow structures oscillating at a single frequency. The algorithm employed here builds upon the SVD of the snapshot data and computes a finite-dimensional approximation of the infinite dimensional Koopman operator~\citep{Tu_etal_2014}.

In the algorithm, the dynamical system (which can be nonlinear) is reconstructed as a linear best-fit (least-squares) approximation $\mathbf{X}' \approx \mathbf{A} \mathbf{X}$, where matrices $\mathbf{X}$ and $\mathbf{X}' \in \mathbb{R}^{m \times n-1}$ store the snapshots in columns. Here, we have $n$ snapshots with $m$ pressure values for different grid points, and matrices $\mathbf{X}$ and $\mathbf{X}'$ are composed by snapshots 1 to $n-1$, and 2 to $n$, respectively.
Then, the best-fit linear operator $\mathbf{A}$ and its eigendecomposition are evaluated to extract the DMD modes $\mathbf{\Phi}$.
The initial amplitudes $b_k$ of each mode are then found such that the solution is approximated by
\begin{equation}
    \mathbf{x}(t) \approx \sum_{k=1}^{r} \mathbf{\phi}_k \exp(\omega_k t) b_k = \mathbf{\Phi} \exp(\mathbf{\Omega} t) \mathbf{b}
    \mbox{ ,}
    \label{eq:DMD}
\end{equation}
where $\omega_k = \ln(\lambda_k)/\Delta t$, and $\mathbf{\Omega}$ is a diagonal matrix whose entries are the $\omega_k$ values. Here, $r$ is the rank of the reduced SVD approximation to $\mathbf{X}$.

In this work, we employ the DMD variation labelled as mrDMD \cite{MrDMD_2016}. The implementation for both algorithms can be found in the PyDMD package \citep{pydmd} available at \url{https://mathlab.github.io/PyDMD/}. 
This algorithm variant consists of a recursive computation of DMD to remove low-frequency, or slowly varying, features from the collection of snapshots~\citep{kutz_et_al}. Its primary advantage stems from its ability to separate long-, medium-, and short-term trends in data. The resulting output, then, provides a means to better analyze transient or intermittent dynamics that the normal DMD fails to capture. Furthermore, the mrDMD is able to handle translational and rotational invariances of low-rank embeddings that are often undermined by SVD-based methods.

In a multiresolution manner, the time domain is divided into two segments recursively to create multiple resolution levels until a desired termination. 
Denoting by $L$, $J$ and $m_l$ the number of decomposition levels, temporal bins per level, and modes extracted at each level $l$, respectively, the dynamical system is expressed as
\[
\mathbf{x}(t) = \sum_{l=1}^{L} \sum_{j=1}^{J} \sum_{k=1}^{m_l} f^{l,j}(t) \mathbf{\phi}_k^{(l,j)} \exp(\omega_k^{(l,j)} t) \, b_k^{(l,j)}
\mbox{.}
\]
In the expansion above, $f^{l,j}(t)$ is an indicator term that acts as a sifting function for each temporal bin, being defined as
\[
f^{l,j}(t) =
\begin{cases}
      1, & t \in [t_j, \, t_{j+1}] \\
      0, & \text{elsewhere}
\end{cases},
\hspace{10pt}
\text{with } j = 1,2,\dots,J \text{ and } J = 2^{(l-1)}.
\]
Artificial high-frequency oscillations may be introduced by the hard cutoff of the time series in each sampling bin but they are naturally filtered out by the lowest frequency selection during the recursion. For a detailed description of the mrDMD algorithm the reader is referred to Ref. \cite{MrDMD_2016}.

\section{Results}\label{sec3}

In the results section, we employ flow modal decompositions to investigate two unsteady flow problems involving transient features and intermittency. In the first, FA-MVEMD and mrDMD are applied to analyze the emergence, development and advection of the dynamic stall vortex and the trailing vortex for a periodic plunging SD7003 airfoil. In the second case, the same techniques are employed to analyze the intermittent and nonlinear phenomena responsible for vortex pairing in a steady NACA0012 airfoil. 

\subsection{Dynamic stall}

A wall-resolved LES is performed for a single plunge cycle of an SD7003 airfoil, for which 8400 snapshots are stored. The airfoil vertical displacement $h(t)$ is specified as a function of non-dimensional time $t$, reduced frequency $k$, and maximum plunge amplitude $h_0$, as $h(t) = h_0 \sin{(2 k t)}$. The maximum plunging amplitude is normalized by the chord length and the airfoil undergoes a variation in effective angle of attack in the range $-6 \degree \leq \alpha_{eff} \leq 22 \degree$. Here, the parameters are set as $k = 0.25$ and $h_0 = 0.5$. Flow conditions are set to freestream Mach number of 0.1 and Reynolds number $6 \times 10^4$. The sampling rate for the acquired snapshots is set to $\Delta t = 0.0015$, yielding a non-dimensional Nyquist frequency of $St = 333$. Here, $St$ refers to the Strouhal number written in terms of the frequency $f$, airfoil chord $L$ and freestream velocity $U_{\infty}$ as $St = f L/U_{\infty}$. This sampling rate is employed to ensure an accurate temporal resolution for the modal decomposition techniques. For more details about the flow simulation, we refer to Ref. \cite{miotto2022analysis}.

\begin{figure}[!h]
\begin{center}
\includegraphics[width=.99\textwidth]{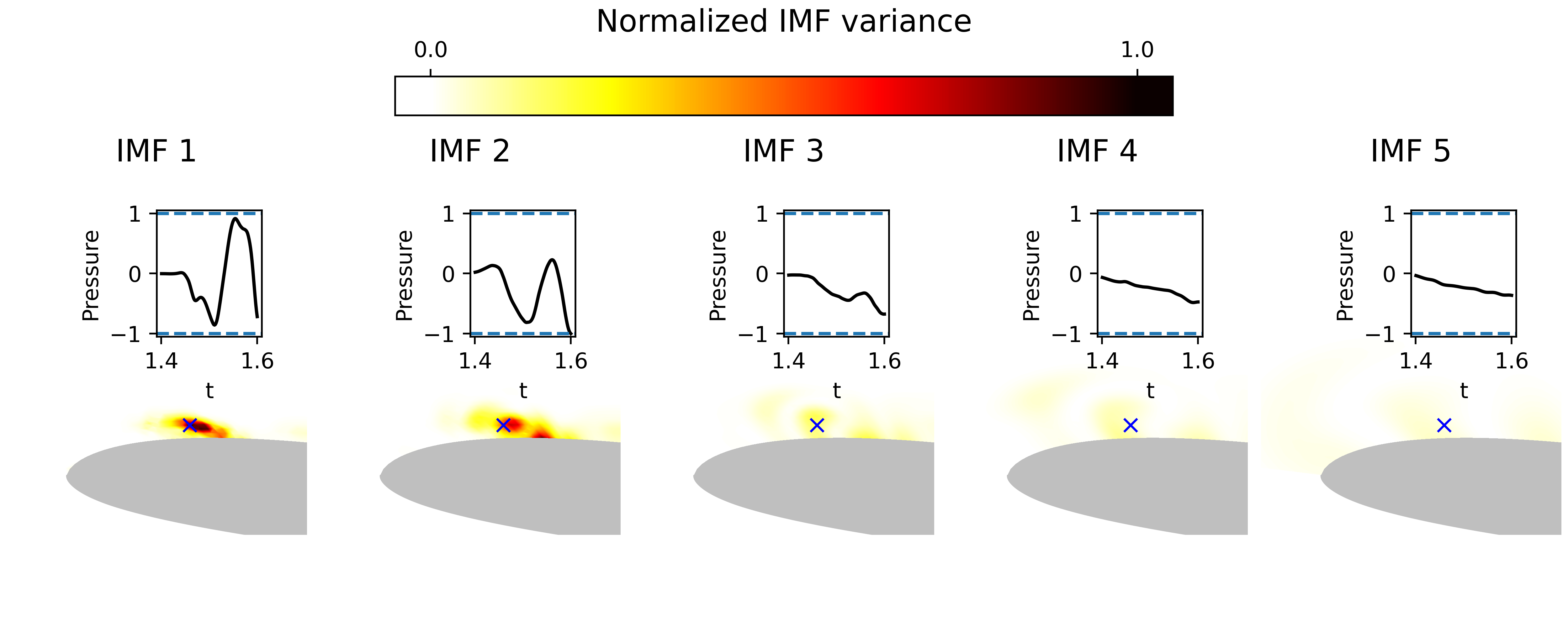}
\end{center}
\caption{Normalized variance contours computed for the time series related to the DSV onset near the airfoil leading edge. From left to right the contour images depict the variances from the IMFs 1 to 5. Inner plots show the IMF signals of pressure variations at the position of maximum variance for the IMF 1, marked by the cross.}
\label{fig : Onset variance}
\end{figure}

The main flow features of different stages of the dynamic stall phenomenon are investigated. First, coherent structures related to the DSV onset are analyzed, followed by its development which occurs after the ejection of small-scale vortical structures. Then, the advection of the DSV and the trailing edge vortex formation are studied. After applying the FA-MVEMD to the pressure dataset, 5 IMFs are extracted. The number of IMFs is a user defined parameter and 5 IMFs was considered to be sufficient to successfully separate the multiple scales of the dynamic stall events as shown in Ref. \cite{Ansell_Karen_2020}. The variance of each IMF during the DSV onset is computed considering the time dependent series of the pressure oscillations at each spatial location resulting in the contours shown in Fig. \ref{fig : Onset variance}. This procedure indicates that the most significant pressure variations in the vicinity of the airfoil leading edge are concentrated in IMFs 1 and 2, with the former presenting the larger variations. Therefore, the following modal analysis results of the DSV onset in terms of the Hilbert spectral analysis are focused on the features depicted in IMF 1, which shows small-scale energetic coherent structures near the wall. This is in line with the analysis from \citet{Visbal_2014}, who observed that large pressure variations near the airfoil wall preceded the onset of dynamic stall for a pitching airfoil at a moderate Reynolds number. Moreover, we found that the instantaneous frequencies obtained for IMF 2 were similar to those from IMF 1, and this ensures that the conclusions would remain unchanged if the former had been employed for the spectral analysis. 

\begin{figure}[!h]
\begin{center}
\includegraphics[width=.495\textwidth]{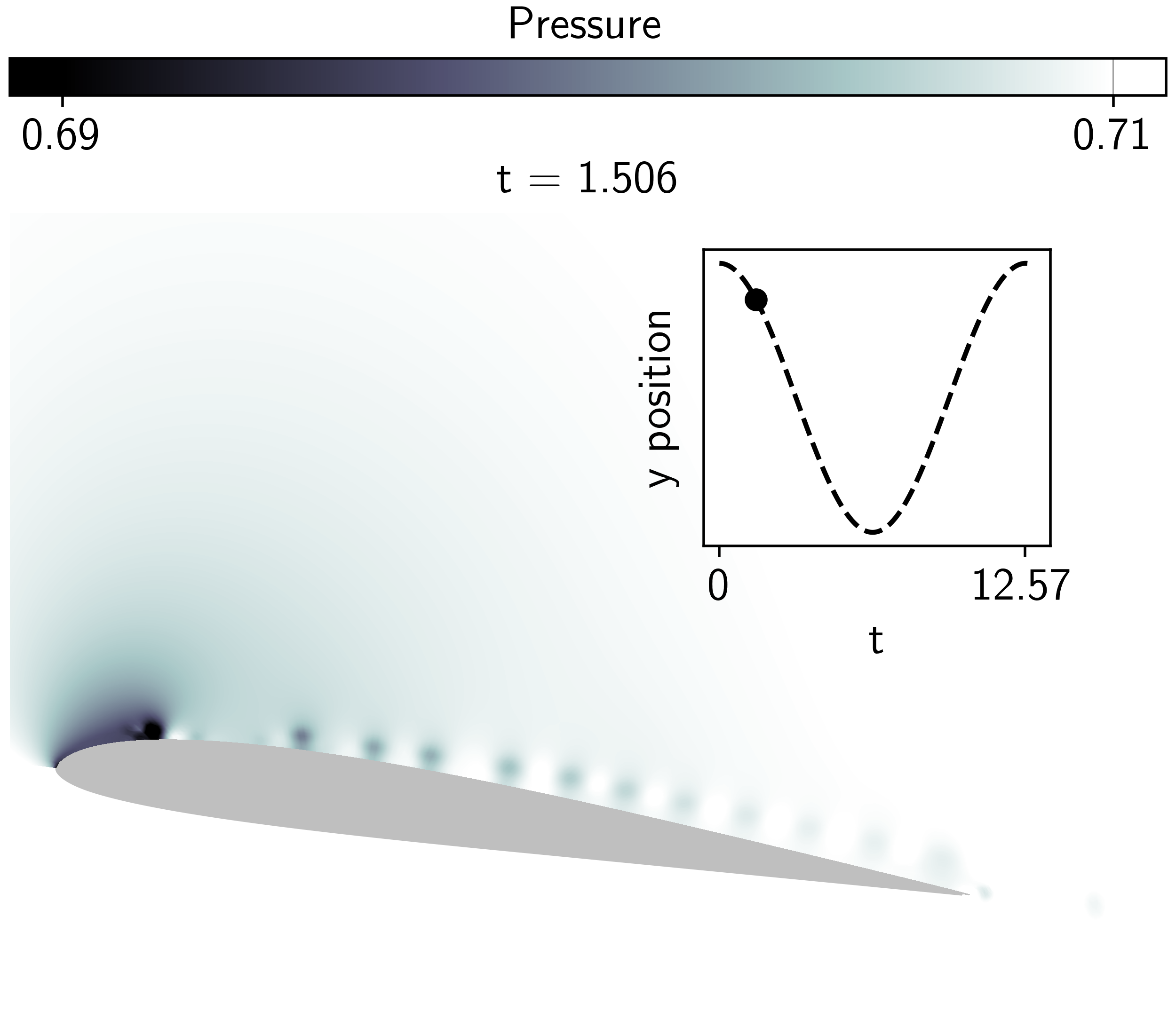}
\includegraphics[width=.495\textwidth]{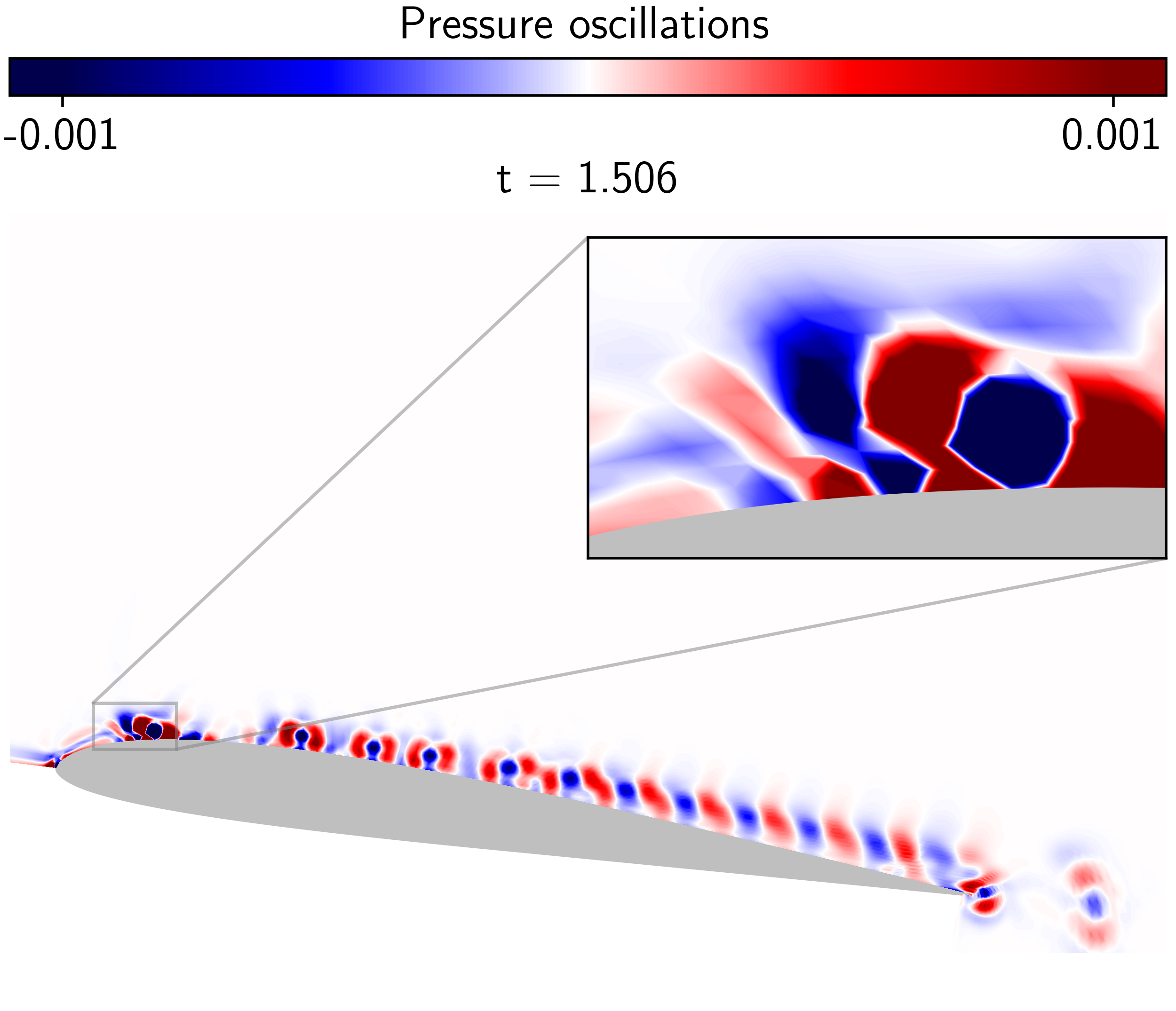}
\includegraphics[width=.495\textwidth]{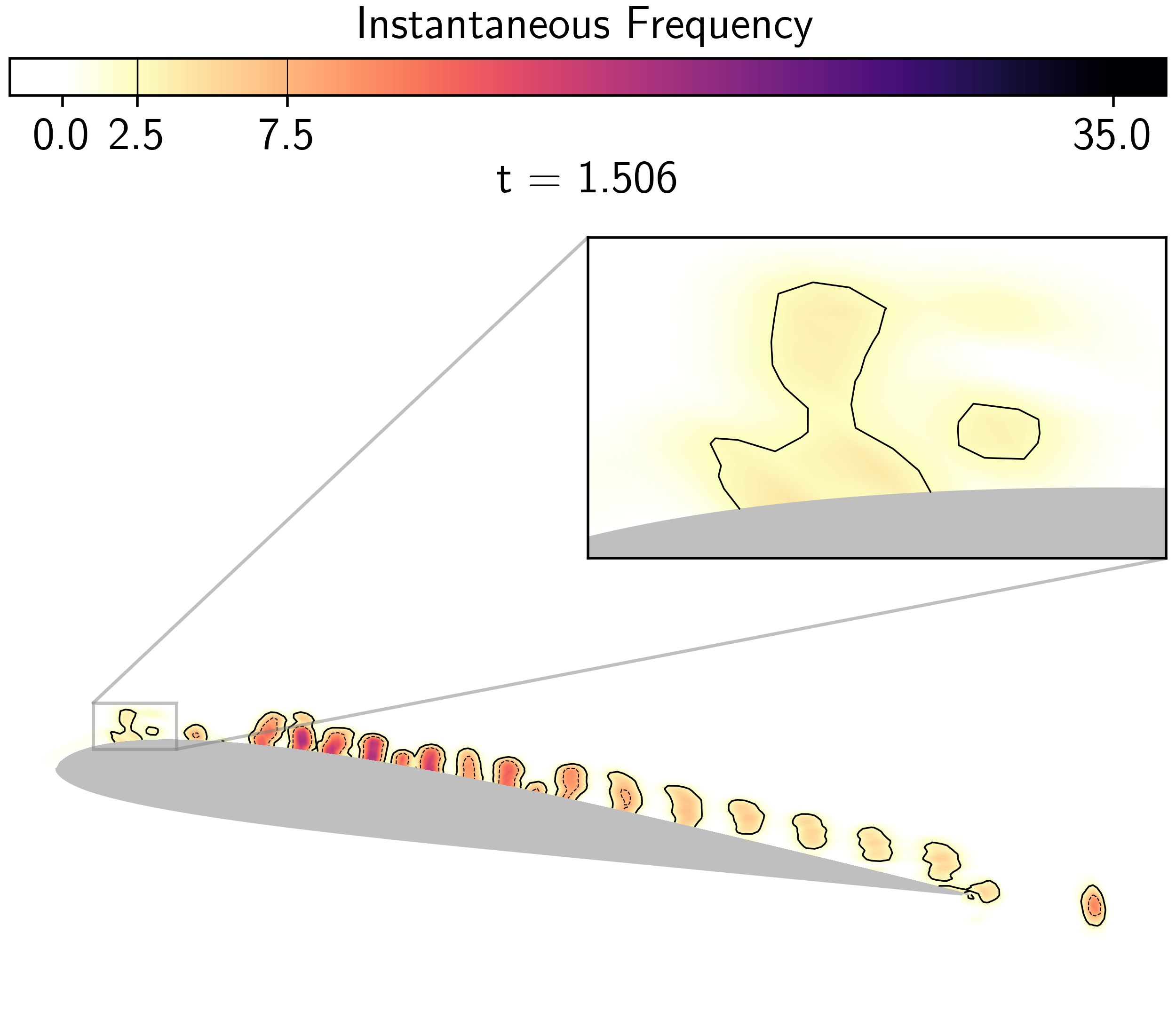}
\includegraphics[width=.495\textwidth]{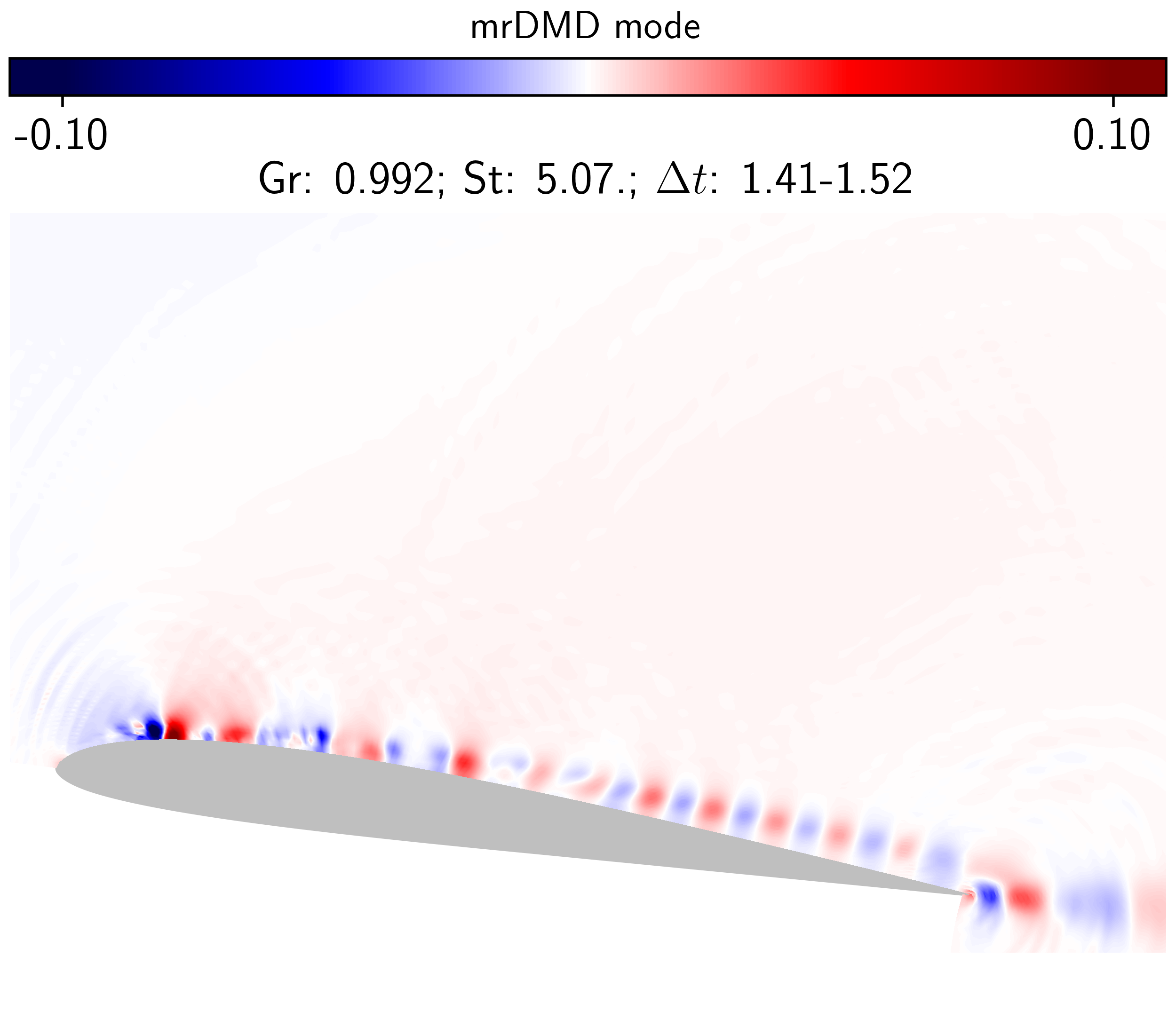}
\end{center}
\caption{Results of modal decomposition prior to the DSV onset. 
In the top images, the left figure shows the instantaneous pressure field from LES, and the right plot shows the pressure oscillations represented by IMF 1. The bottom left and right figures present, respectively, the instantaneous frequencies from the Hilbert spectral analysis and a mrDMD mode corresponding to a temporal window and Strouhal number similar to those from the IMF.}
\label{fig : Onset}
\end{figure}

Figure \ref{fig : Onset} presents contours of the instantaneous nondimensional pressure field (top left), the pressure distribution from IMF 1 (top right), the instantaneous frequency computed by the Hilbert transform of the same IMF (bottom left), and a selected mrDMD mode with a frequency similar to that mostly observed in the Hilbert transform (bottom right). The contours are computed for the instant when a Kelvin-Helmholtz instability reaches the low pressure zone at the leading edge, as the flow rapidly accelerates in this region during the beginning of the airfoil descending trajectory. In the top right corner of the first image, a sinusoidal curve is drawn in the inset to represent the vertical position prescribed by the airfoil over time, over which the black dot indicates the airfoil position for the present snapshot. It is worth noting that a full cycle of the airfoil motion has a non-dimensional period of $\pi / k \approx 12.57$.
We can see that the IMF 1 from the EMD is able to identify the coherent structure that emerges closer to the leading edge as well as the train of Kelvin-Helmholtz instabilities. The IMF depicts the instantaneous vortical structures by the blue contours, which represent negative pressure values. 

For the mrDMD mode (bottom right image), the time window in which the mode exists is indicated in the plot, along with its growth rate ($Gr$) and Strouhal number ($St$). The time under consideration for the IMF 1, $t = 1.506$, is within the time window of the mrDMD bin. 
The mrDMD mode presents the vortices of the Kelvin-Helmholtz instability located upstream the mid-chord in a more distorted way compared to the IMF. This is due to the fact that these vortices are being advected at a higher speed, as can be seen in the instantaneous frequency contours in the lower left image. The Hilbert transform shows that the upstream structures have higher frequencies than those from the mrDMD mode. 

In a previous study of the same plunging airfoil  \cite{ramos2019active}, it was shown that a leading-edge flow actuation at
at a frequency range of $ St = 2.5-7.5 $ could disrupt the formation of the DSV. In the instantaneous frequency contours from the Hilbert spectral analysis, black lines are included to indicate the regions where the flow structures present frequencies within this range. For the structure originating at the leading edge, the instantaneous frequencies  point to a value within this particular range, which may justify the success of the flow actuation. In the figure, black dashed lines are also shown to delimit the regions where the instantaneous frequencies are above $St = 7.5$. 

The bottom right image corresponds to a single mrDMD mode obtained to represent the system dynamics in that particular time window. However, a complete mrDMD decomposition of the structure represented in the figure should be obtained by considering other modes at different levels spanning the same time interval. In this case, a similar band of frequencies as those displayed by the Hilbert spectral analysis would be found. Hence, it can be said that both methods are successful in decomposing and identifying the frequencies related to the particular flow event, but the combination of EMD with the Hilbert spectral analysis is preferred, as it provides a direct correlation between the frequency content and the instantaneous structures in the flow in a more straightforward way. This is true since a single IMF can be used to obtain the instantaneous frequencies of the flow structures, while the mrDMD requires a search for particular frequencies at the various bins in the same temporal window.
For example, at the instant analyzed, the flow depicts the formation of a leading edge coherent structure, the evolution of a train of Kelvin-Helmholtz vortices, and the shedding of a trailing edge vortex, which can be seen in the IMF 1. 
Each of these events have their own timescales, and the decomposition performed by the mrDMD is not able to pin the frequencies directly to the isolated structures as in the Hilbert spectral analysis. Also, the number of IMFs produced by the EMD is far less than the number of mrDMD modes, making the combination of EMD with Hilbert spectral analysis more convenient to identify flow structures of interest. 
\begin{figure}[htb!]
\begin{center}
\includegraphics[width=.495\textwidth]{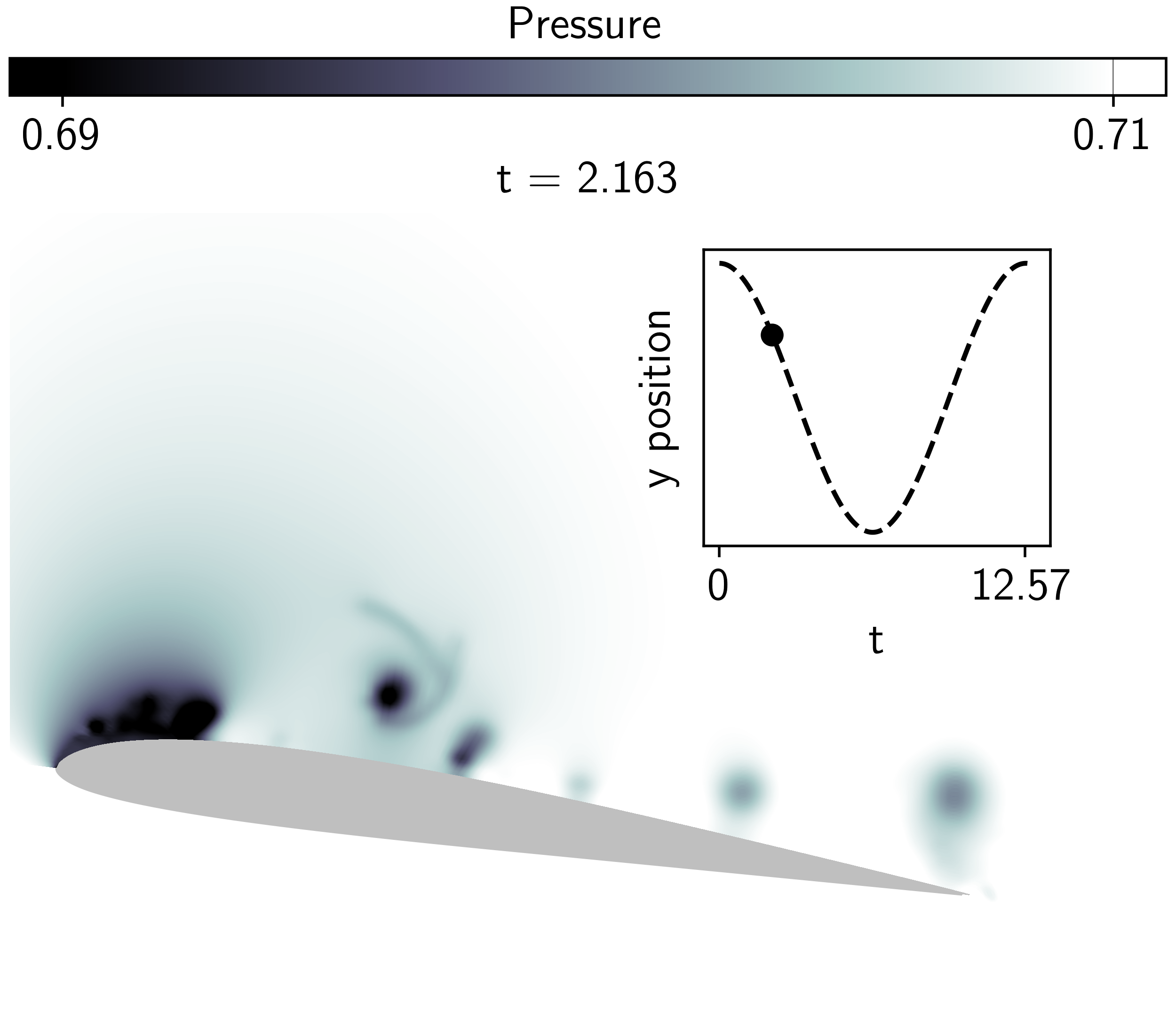}
\includegraphics[width=.495\textwidth]{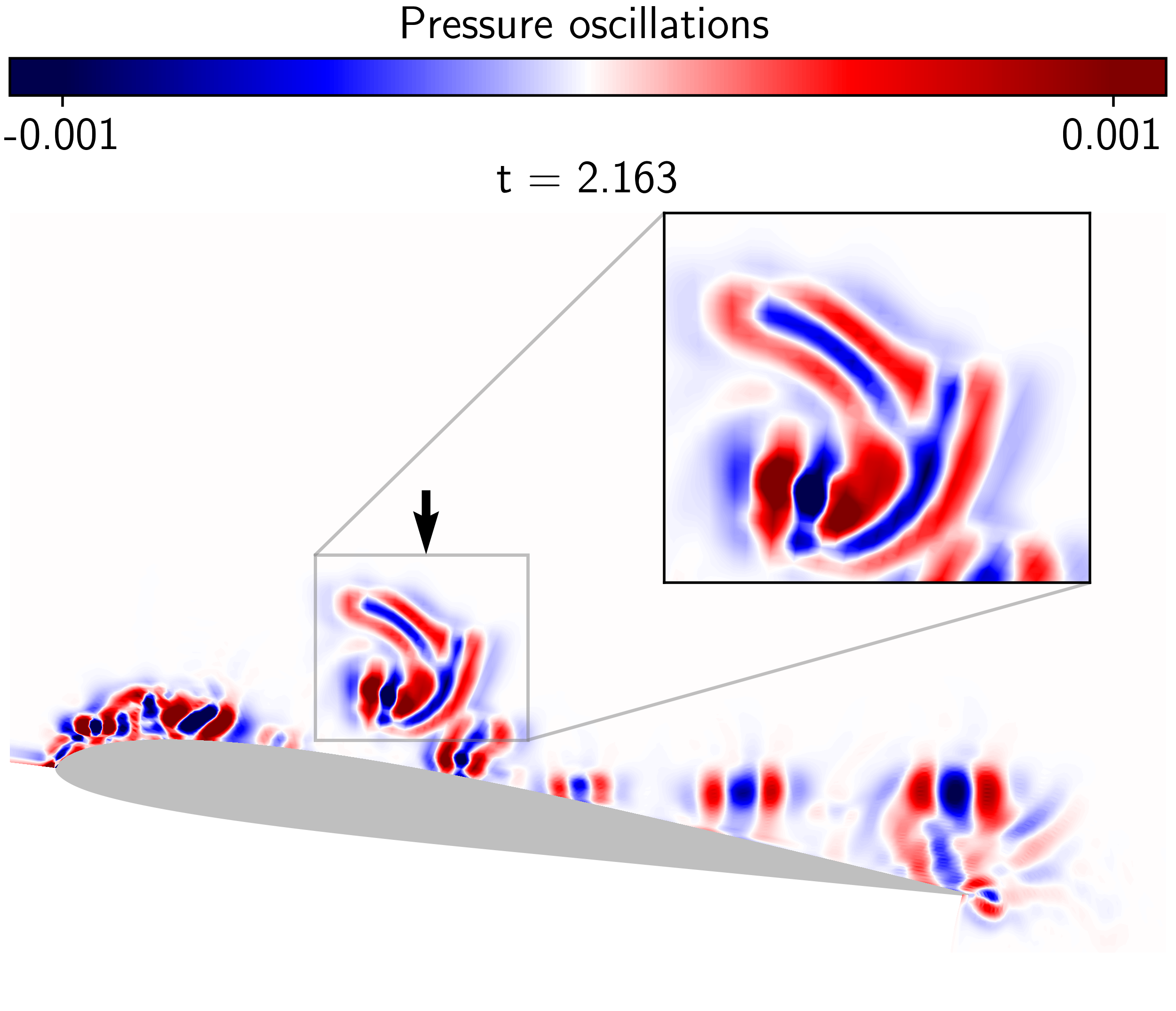}
\includegraphics[width=.495\textwidth]{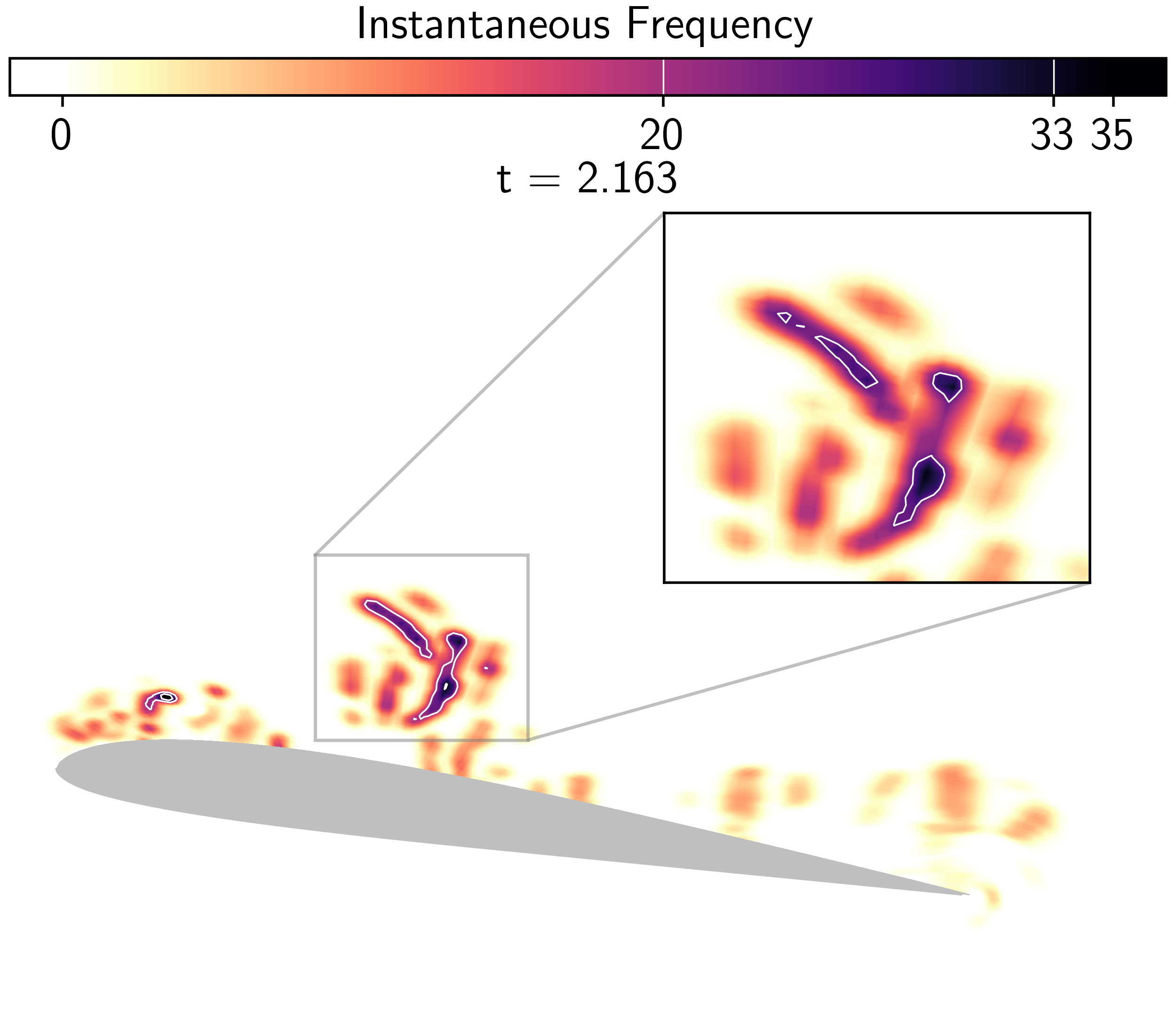}
\includegraphics[ width=.495\textwidth]{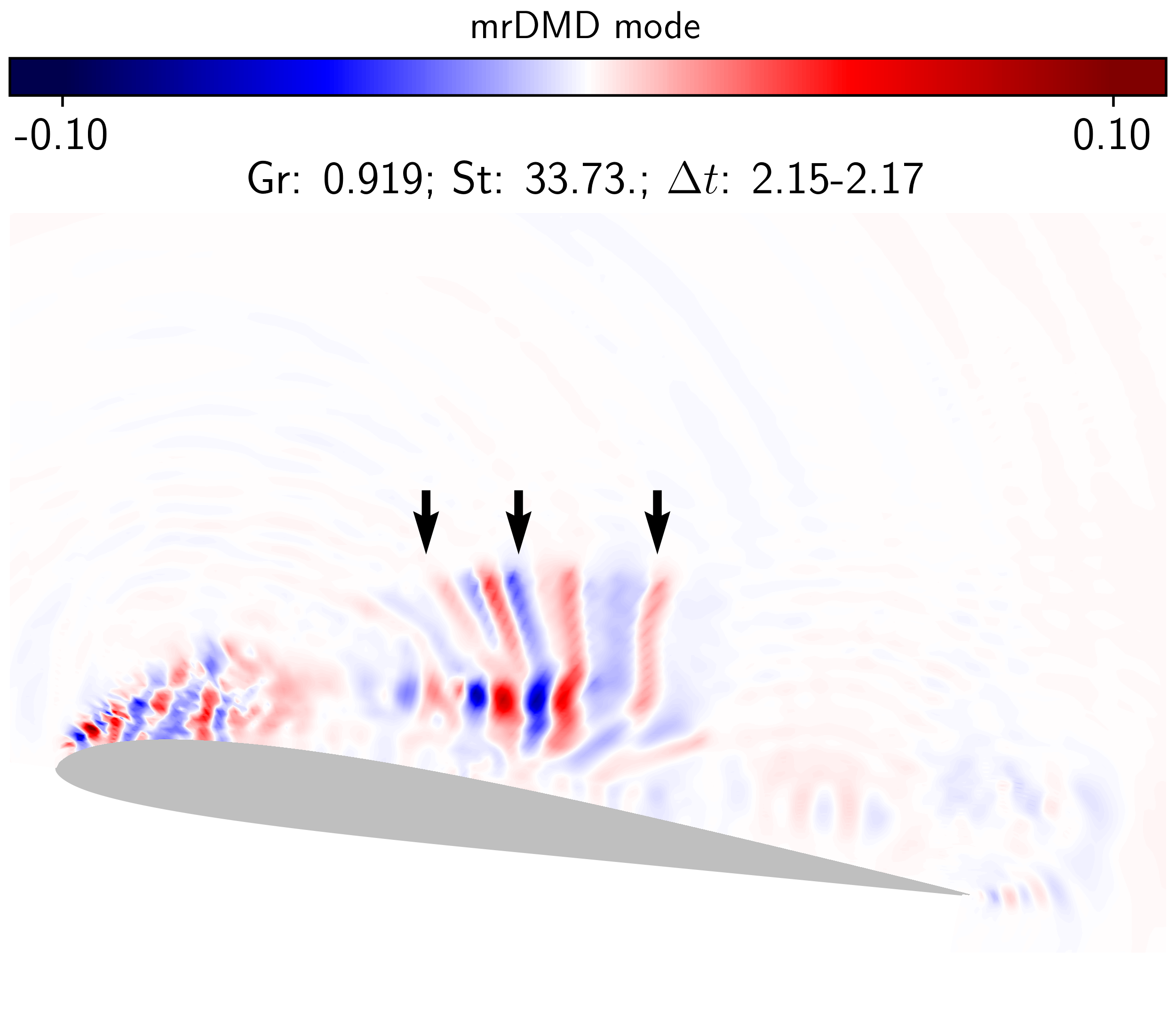}
\end{center}
\caption{Results from modal decomposition during the early development stages of the DSV. In the top images, the left figure shows the instantaneous pressure field from LES, and the right one shows the pressure oscillations represented by IMF 1.
The bottom left and right figures present, respectively, the IMF 1 instantaneous frequencies from the Hilbert spectral analysis, and a mrDMD mode corresponding to a temporal window and Strouhal number similar to that from the IMF.}
\label{fig : DSV development}
\end{figure}

Following the airfoil downward motion, a coherent structure grows being fed by vorticity that accumulates at the leading edge. This is illustrated in Fig. \ref{fig : DSV development}, which contains the instantaneous pressure (top left), the IMF 1 (top right) and its correspondent Hilbert spectrum (bottom left), and a mrDMD mode related to the equivalent time window (bottom right). The IMF highlights the fact that the DSV is formed by a coalescence of multiple small-scale vortices emerging from the leading edge during the airfoil descending stage. During the process of vorticity accumulation at the leading edge, some structures eject and are transported with the flow as can be also seen in the figure. For example, the structure marked by the black arrow in the IMF plot consists of a vortex and a tail, a process which occurs due to the merging of vortices with different sizes as described by \citet{Dritschel_1992}. The IMF 1 is able to represent these instantaneous features by the negative pressure fluctuations with blue contours.
Both processes of the DSV development and the advection of ejected vortices are governed by high-frequency dynamics, as demonstrated by the Hilbert spectral analysis and the mrDMD mode with an equivalent frequency. On the other hand, the vortices near the trailing edge have lower instantaneous frequencies, as indicated by the Hilbert transform. For this reason, they are represented in lower levels of the mrDMD decomposition and do not appear in this particular mode. Hence, the contours obtained by the Hilbert spectral analysis are able to condense a larger amount of information in a single IMF.

Although both modal decomposition techniques are able to identify regions with dynamics at a similar frequency range, the mrDMD shows some limitations. For instance, in Fig. \ref{fig : DSV development}, while the spatial support of the IMF is more compact and better localized, the equivalent mrDMD spatial mode represents the same coherent structure in a scattered fashion. Precisely, we observe that the mid-chord vortex and its upward-pointing tail, are dispersed in the mrDMD mode as indicated by the vertical arrows. This distortion arises from the limitations of SVD-based methods to capture translational and rotational features as previously discussed. In fact, the beginning and the end of the distorted region coincide with the trajectory of the flow structure in the time interval of the mrDMD bin. The multi-resolution approach reduces the problems with translational invariances, typical of advection-dominated flows, for the complete reconstruction of the original data. However, the individual modes obtained by the recursive application of the standard DMD in the temporal bins still does not completely eliminate this unwanted effect.
As a consequence, the spatial characterization of transient coherent structures may be compromised. In fact, when using high sampling rates to form the DMD data matrix, we observed significant errors in the reconstructed flow dynamics from the mrDMD modes at moments when the flow presents translation or rotational features (not shown here for brevity). On the other hand, the IMFs generated by the EMD do no suffer from distortion effects as can be verified by comparing the IMF 1 with the original pressure data.
\begin{figure}[!h]
\begin{center}
\includegraphics[width=.495\textwidth]{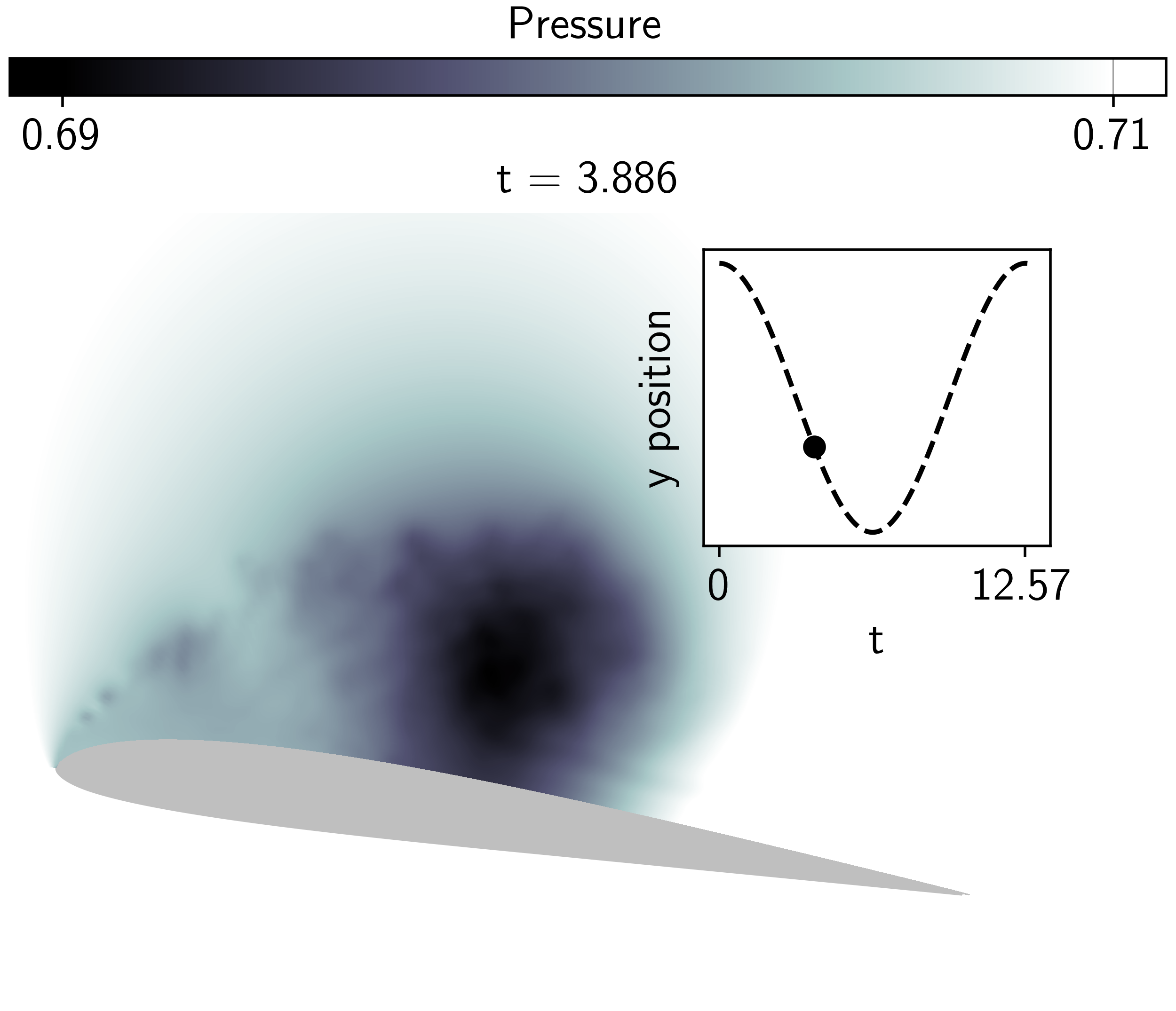}
\includegraphics[width=.495\textwidth]{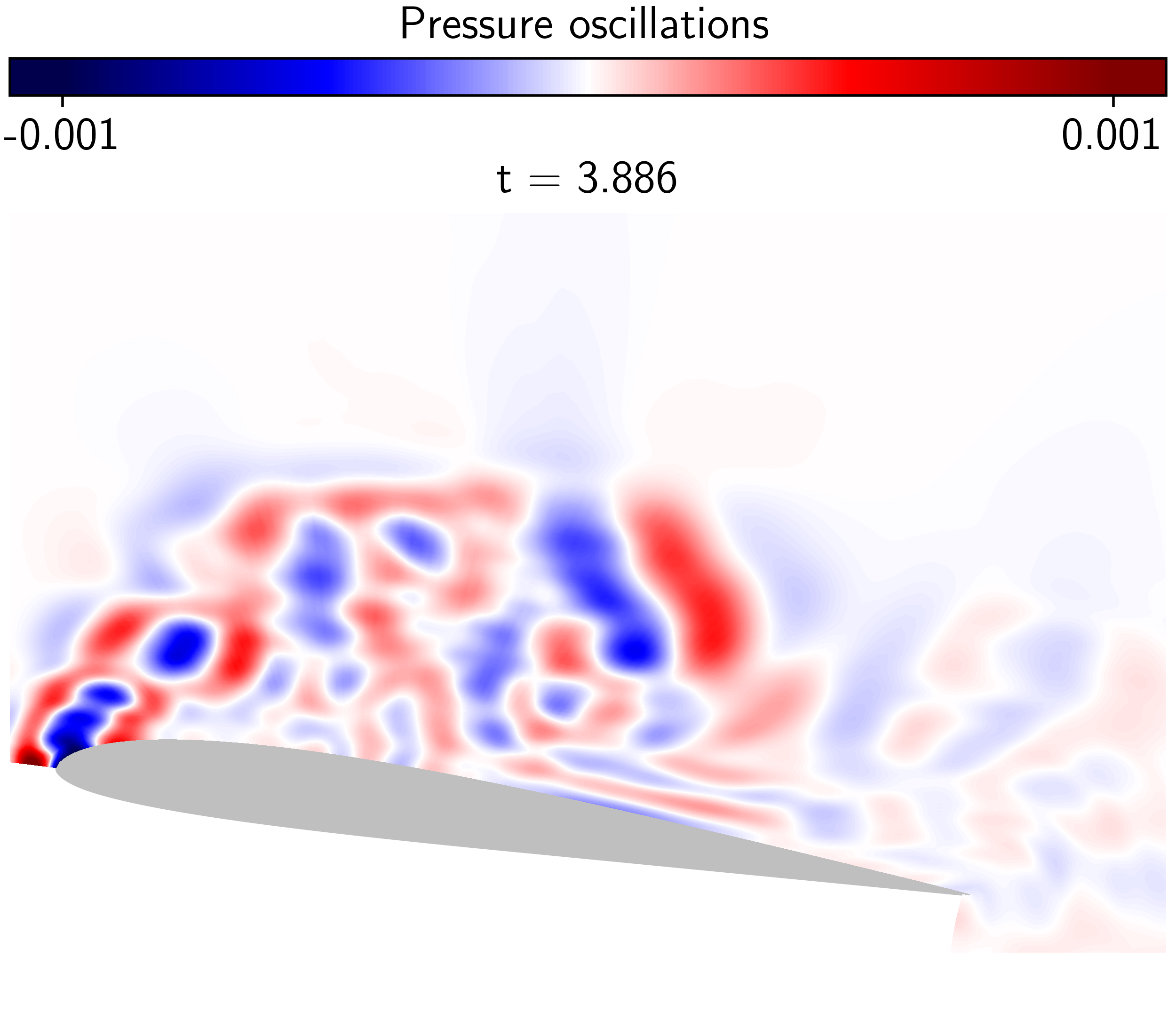}
\includegraphics[width=.495\textwidth]{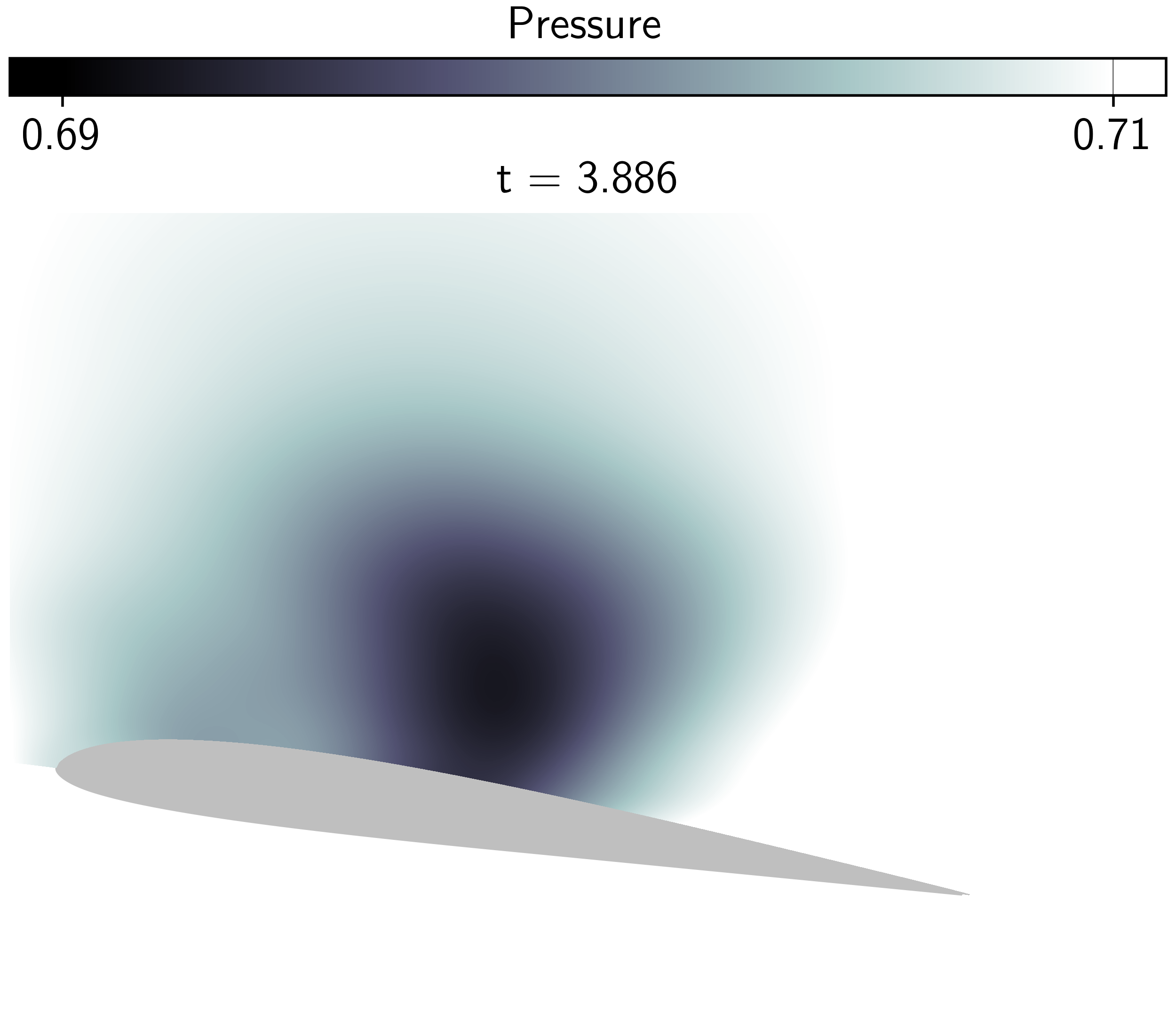}
\includegraphics[width=.495\textwidth]{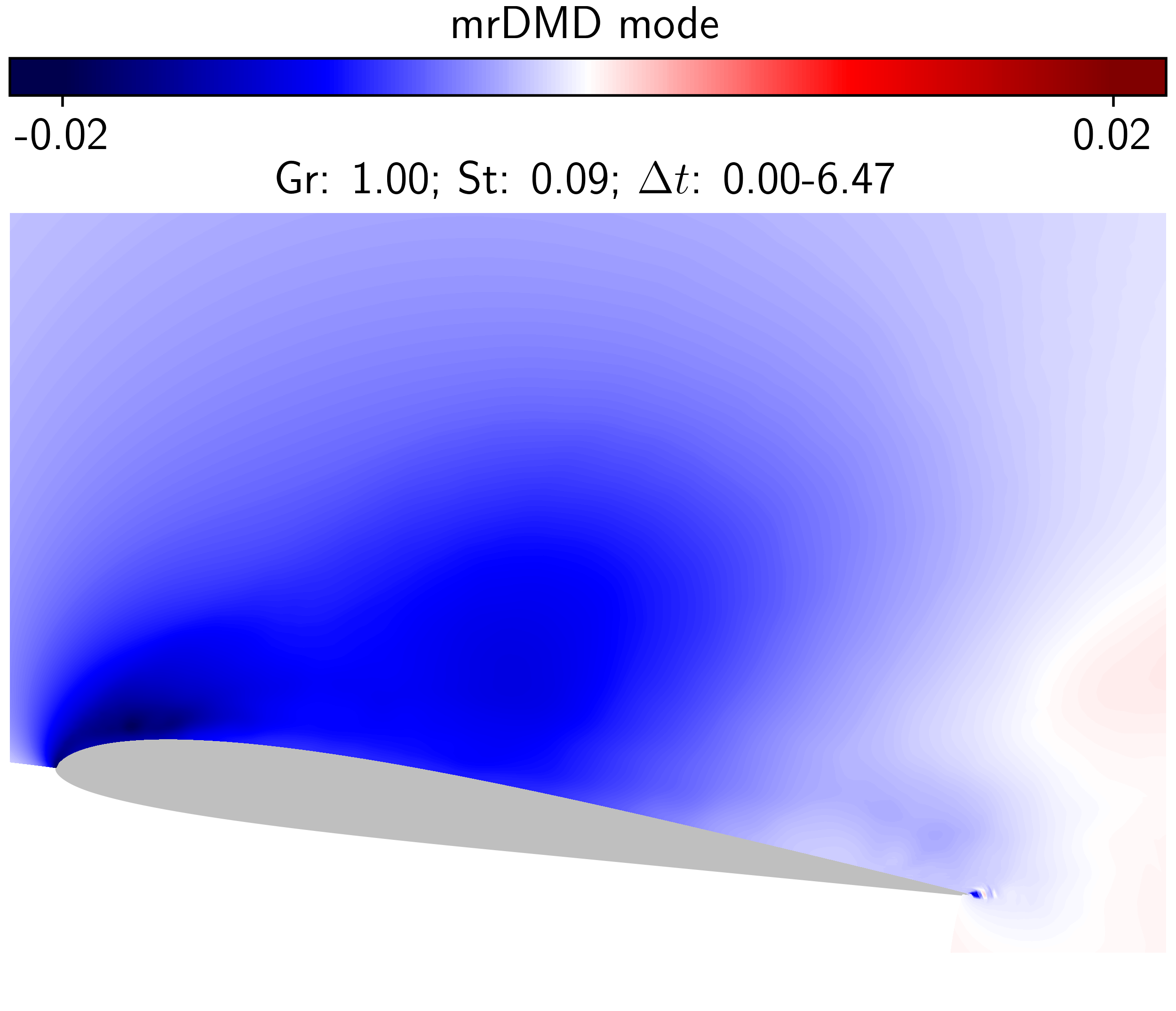}
\end{center}
\caption{Results from modal decomposition of the DSV transport. In the top images, the left figure shows the instantaneous pressure field from LES, and the right one shows the pressure oscillations represented by IMF 3.
The bottom left and right figures present, respectively, the EMD residue and a mrDMD mode corresponding to a temporal window similar to that from the IMF.}
\label{fig : DSV}
\end{figure}

\begin{figure}[b!]
\begin{center}
\includegraphics[width=.495\textwidth]{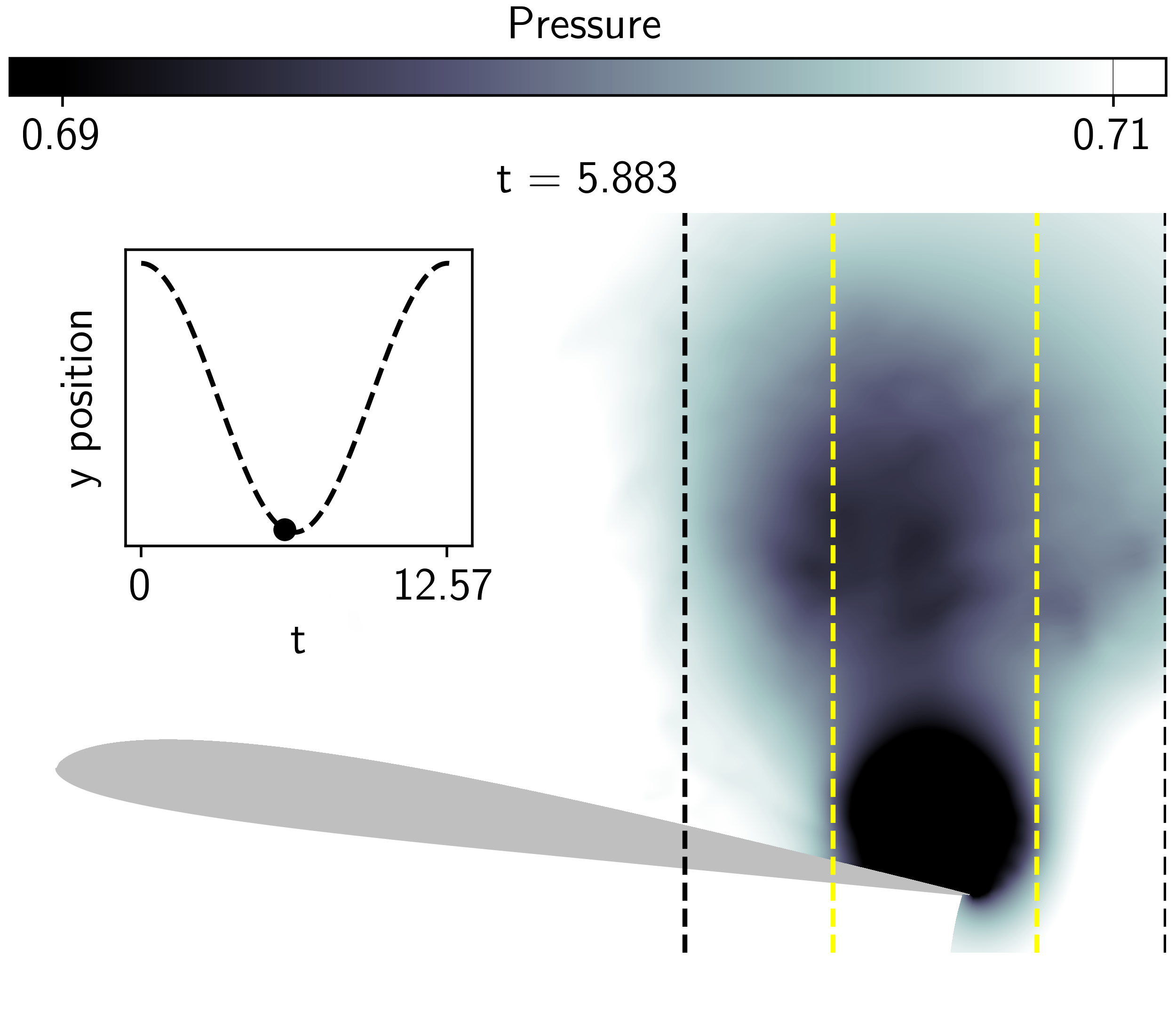}
\includegraphics[width=.495\textwidth]{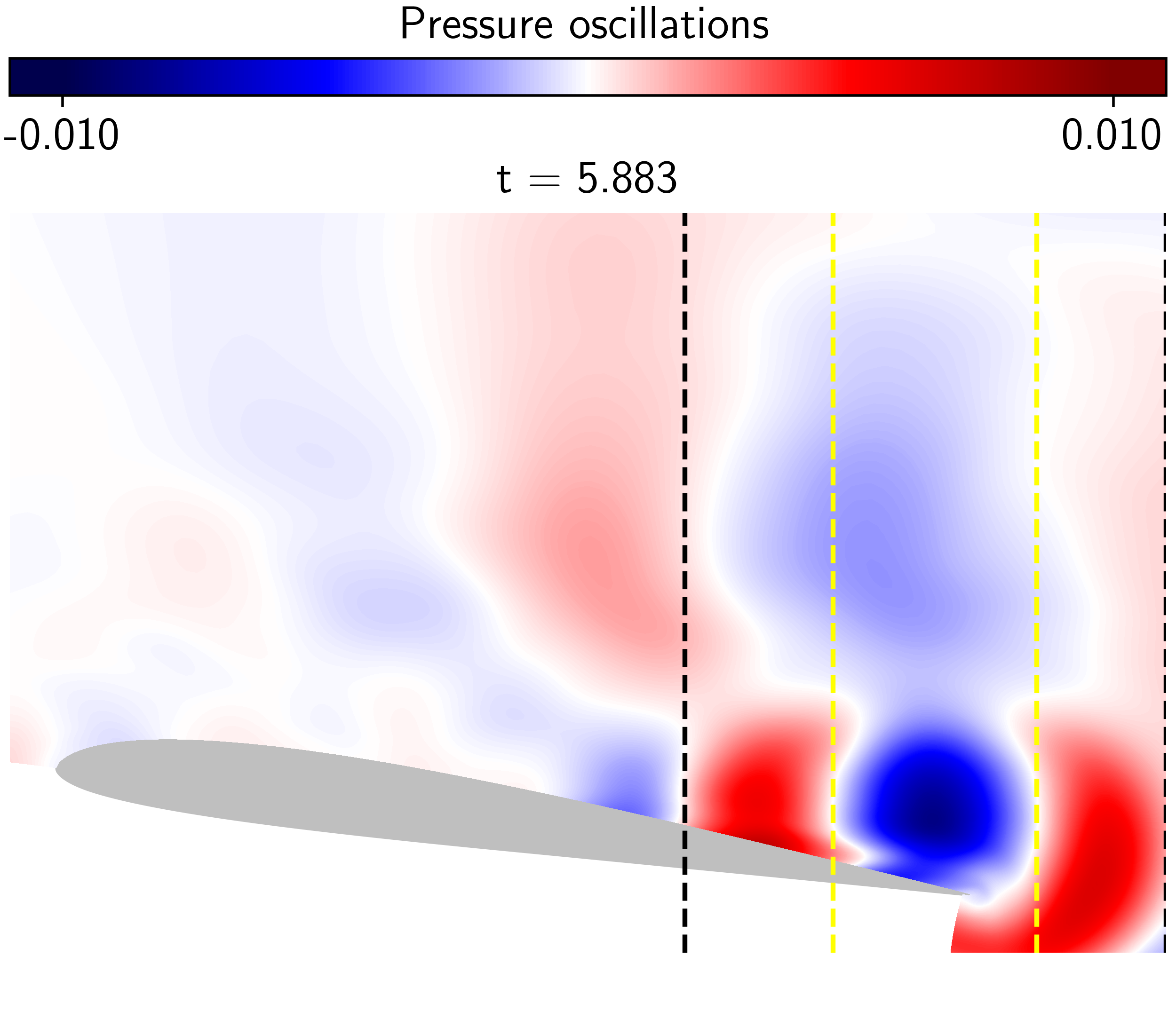}
\includegraphics[width=.495\textwidth]{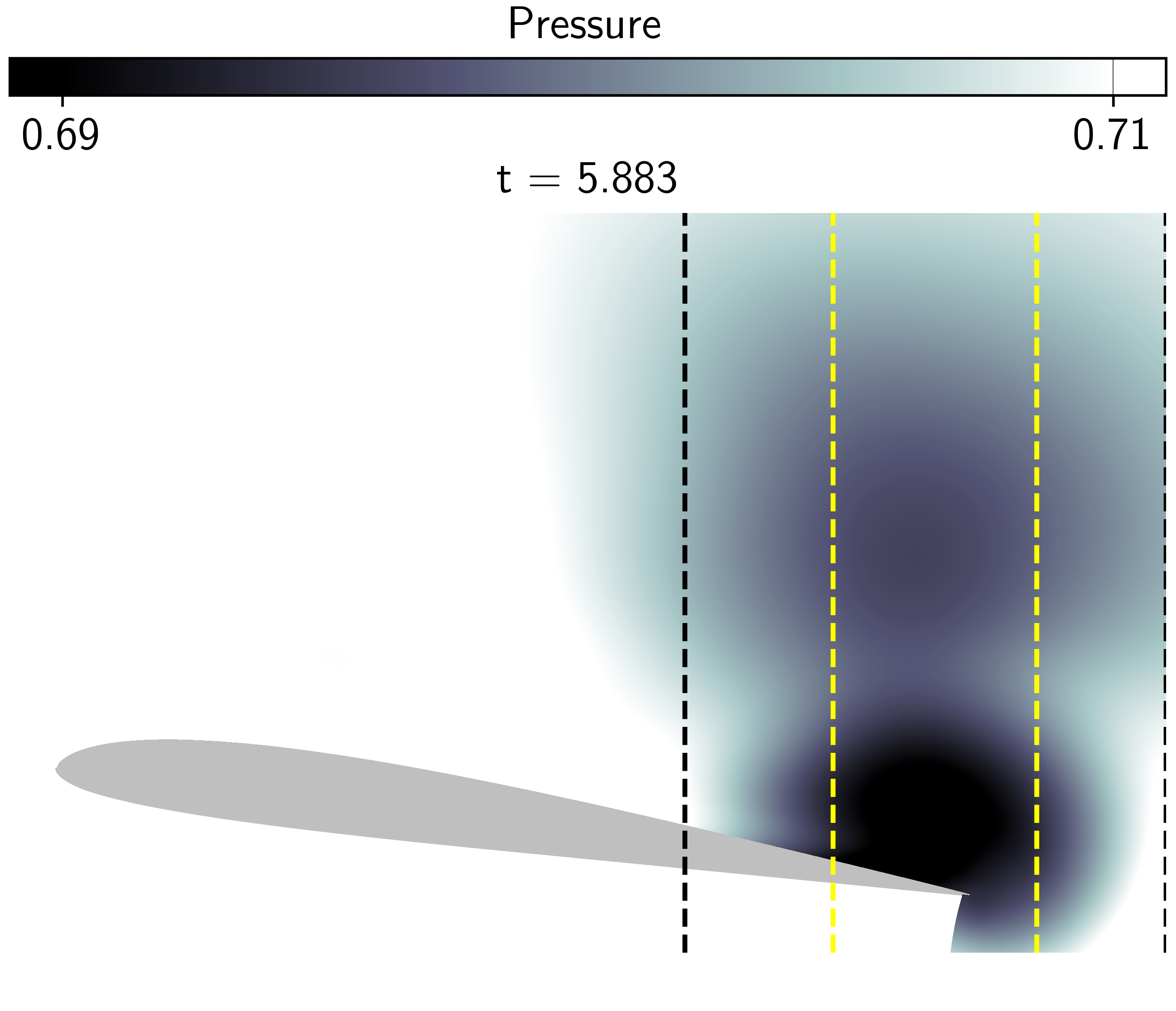}
\includegraphics[ width=.495\textwidth]{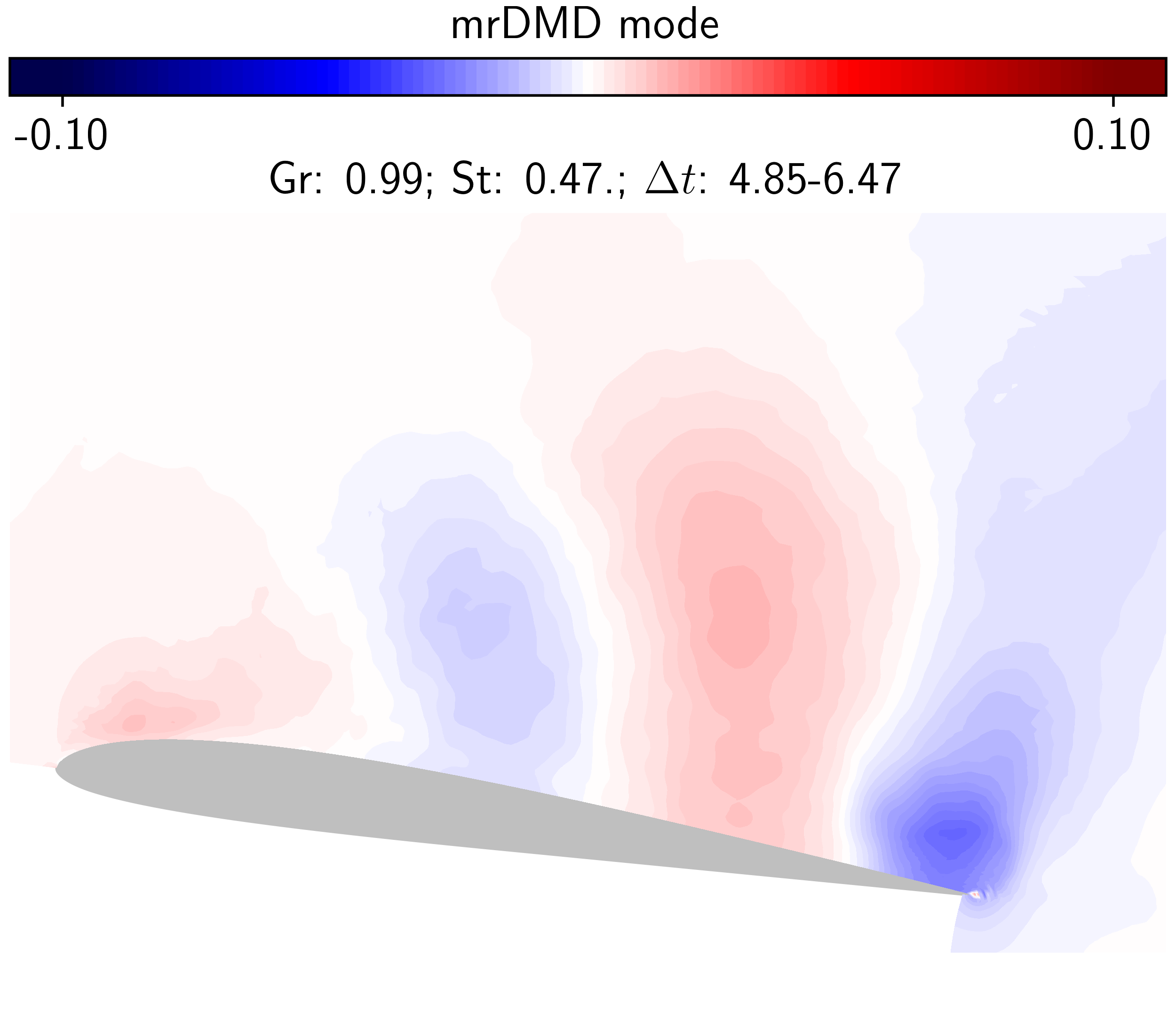}
\end{center}
\caption{Results from modal decomposition of the TEV. In the top images, the left contour shows the instantaneous pressure field from LES, and the right plot shows the pressure oscillations represented by IMF5.
The bottom left and right figures present, respectively, the EMD residue and a mrDMD mode corresponding to a temporal window similar to that from the IMF.}
\label{fig : TEV}
\end{figure}

After the mid-chord vortices are shed along the wake, the main flow feature becomes the transport of the DSV, as illustrated in Fig. \ref{fig : DSV}. At the instant displayed by the instantaneous pressure contours in the top left image, a feeding vortex sheet still connects the DSV to the leading edge. In the figure, the IMF 3 is also presented for the same snapshot depicting small-scale coherent structures along the DSV. By definition, the IMFs contain the local flow pressure oscillations while the EMD residue shows the non-stationary trends of the flow. For this transient problem, the DSV is better characterized by the residue which presents itself as a single low pressure structure that is shown centered at the mid-chord in the bottom left image. The absence of the feeding sheet in this residue is evident from the comparison with the original pressure field. In the mrDMD mode, on the other hand, the DSV signature is present in the lower levels of the multi-resolution decomposition due to the fact that its advection process is slower than the small-scale local events of the dynamic stall. The mrDMD mode shown in Fig. \ref{fig : DSV} exemplifies this observation as it represents the advection of the DSV associated with a low frequency. As can also be observed from the figure, this mode also shows some contamination from the TEV and the processes related to the DSV formation, taking place at the leading edge. On the other hand, high frequency DMD modes (not shown here) present similar features compared to the IMFs, such as the small scale structures within the DSV displayed in the IMF 3 plot above. This suggests that the coalescence that gives rise to the DSV in the early stages of the flow is a continuous process, with small-scale vortices shedding from the airfoil leading edge and interacting with each other within the DSV. Indeed, the 3D Q-criterion contours shown by Ref. \cite{ramos2019active} indicate that the DSV is composed by a broad range of turbulent scales for the present flow configuration.


Finally, the transport of the DSV changes the circulation along the airfoil and induces the formation of a trailing edge vortex. This latter structure initially moves upstream towards the airfoil, and causes the DSV to detach from the surface upon reaching the trailing edge region, as illustrated in Fig. \ref{fig : TEV}. In this figure, the contours of IMF 5 are represented in the upper right corner and the EMD residue is shown in the lower left corner. In the figure, yellow lines mark the region delimited by the approximate diameter of the TEV. The IMF 5 and the residue display both the TEV and the detached DSV. From the present results, it is clear that the IMFs generated by the EMD capture the instantaneous negative pressure oscillations induced by the vortices, but also display spurious extrema to satisfy the number of zero-crossings which are intrinsic to the IMFs. This effect leads to positive pressure fluctuations which are delimited by black lines adjacent to the TEV, shown in the IMF 5. Such oscillations create a small distortion in the TEV pressure signature shown in the residue. This effect, which compromises the spatial representation of structures in the EMD, is related to the impulse response property of noise-assisted EMD methods \cite{Wu_2009, Rehman_2013}. We point out that these oscillations are non-physical, being a mathematical artifact of the method to obtain a well-behaved Hilbert transform.
The mrDMD mode shown in the lower right corner also captures the TEV. Nevertheless, as the large-scale low-frequency structures are represented at the lower levels of the multi-resolution approach, a long time window must be employed in the decomposition. Therefore, the mrDMD modes that represent these structures are compromised by the same effects as the original DMD algorithm, i.e., the method assumes that the coherent structures are temporally oscillating along the entire temporal window. This is evident in the mode displayed in the figure, which represents the TEV together with the pressure signature of the DSV advection. However, the TEV arises only at later stages of the airfoil motion, while the DSV is transported during the initial instants of the time window corresponding to this DMD mode.

\subsection{Transitional airfoil flow}

\begin{figure}[!h]
\begin{center}
\includegraphics[width=.95\textwidth]{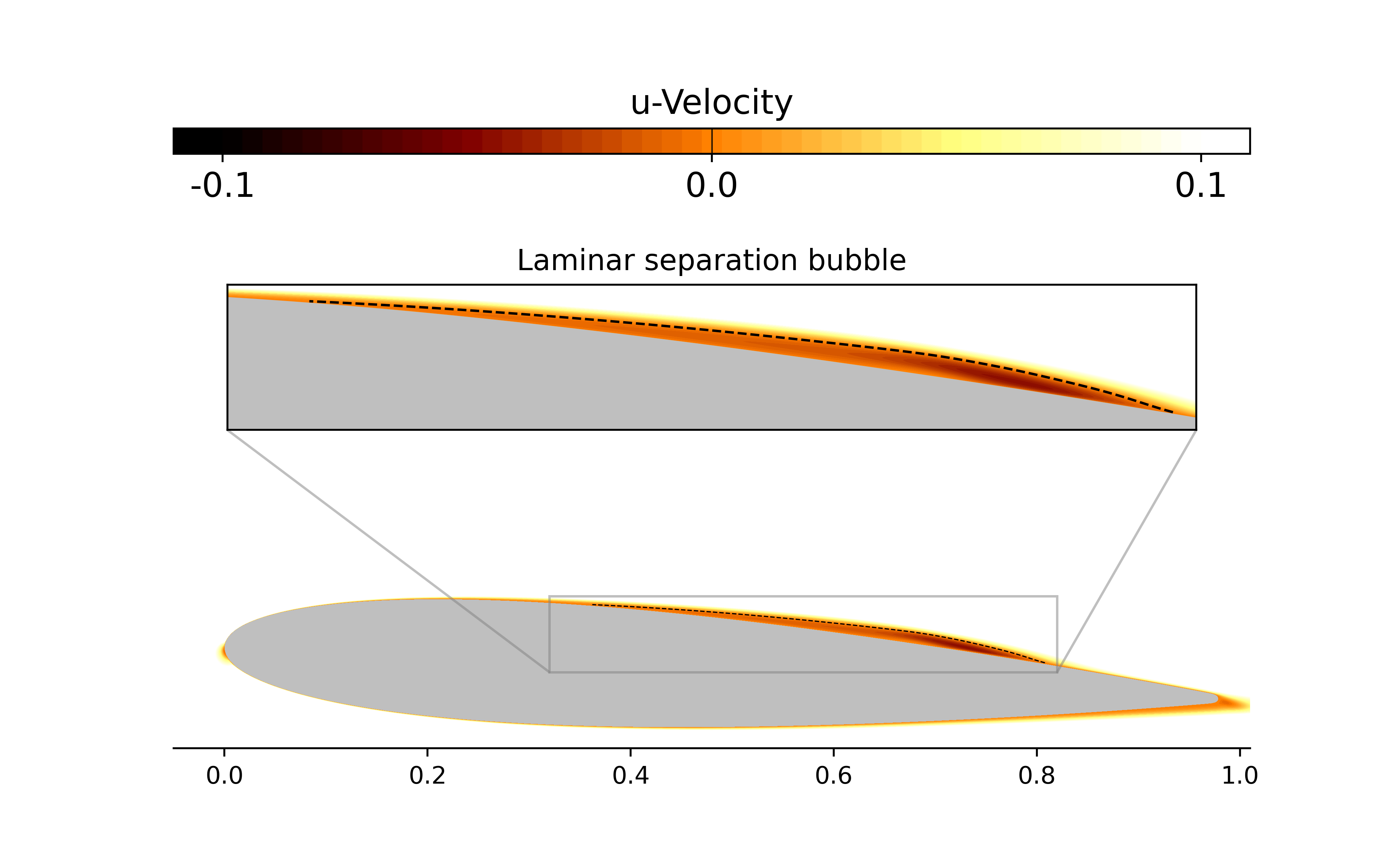}
\end{center}
\caption{Time averaged u-velocity contours for the NACA0012 airfoil with 3 deg. angle of attack. The presence of a laminar separation bubble in the airfoil suction side is highlighted by the black dashed line delimiting the reversed flow region.}
\label{fig : laminar separation bubble}
\end{figure}

In this problem, modal decomposition is applied to the transitional flow over a NACA0012 airfoil with an angle of attack of 3 deg, freestream Mach number 0.3 and
Reynolds number $5 \times 10^4$. The wall-resolved LES dataset from \citet{ricciardi_wolf_taira_2022} is analyzed through the FA-MVEMD and mrDMD. As depicted in Fig. \ref{fig : laminar separation bubble}, in this flow, a laminar separation bubble is observed over the airfoil suction side and it leads to vortex shedding that advects towards the trailing edge causing tonal noise generation. Intermittent vortex interaction occurs due to bubble flapping, which in turn leads to a frequency modulation of the coherent structures, resulting in vortex merging or bursting of turbulent packets. This mechanism also modulates the amplitude of the incident pressure signal that leads to different levels of noise emission at the trailing edge. In Ref. \cite{ricciardi_wolf_taira_2022}, spectral analysis of the LES results and biglobal stability analysis revealed a set of discrete frequencies of the coherent structures at different chord positions. According to the previous authors, the main frequencies span a range of Strouhal numbers $3 \leq St \leq 6$ with a $\Delta St = 0.5$. Here, flow modal decomposition of this intermittent flow is applied to further elucidate the vortex pairing mechanism on the airfoil suction side.

To investigate the intermittent vortex pairing, EMD and mrDMD are applied to a set of 5000 spanwise-averaged pressure snapshots obtained from the LES. This is justified since we are interested in analyzing two-dimensional coherent structures which are responsible for efficient trailing edge noise generation. In this section, we focus on the presentation of the results from the FA-MVEMD analysis because it effectively enhances our understanding of the vortex pairing mechanism occurring on the suction side of the airfoil. Next, the mrDMD modes corresponding to the time intervals employed in the EMD analysis are presented, followed by a comparison highlighting the disparities in their ability to offer meaningful physical insights. The data used for the modal decomposition correspond to approximately 20 dimensionless convective time units. The sampling rate for the acquired snapshots is set to $ \Delta t = 0.00375$ yielding a non-dimensional Nyquist frequency of $St = 133$.

\begin{figure}[!h]
\begin{center}
\includegraphics[width=.99\textwidth]{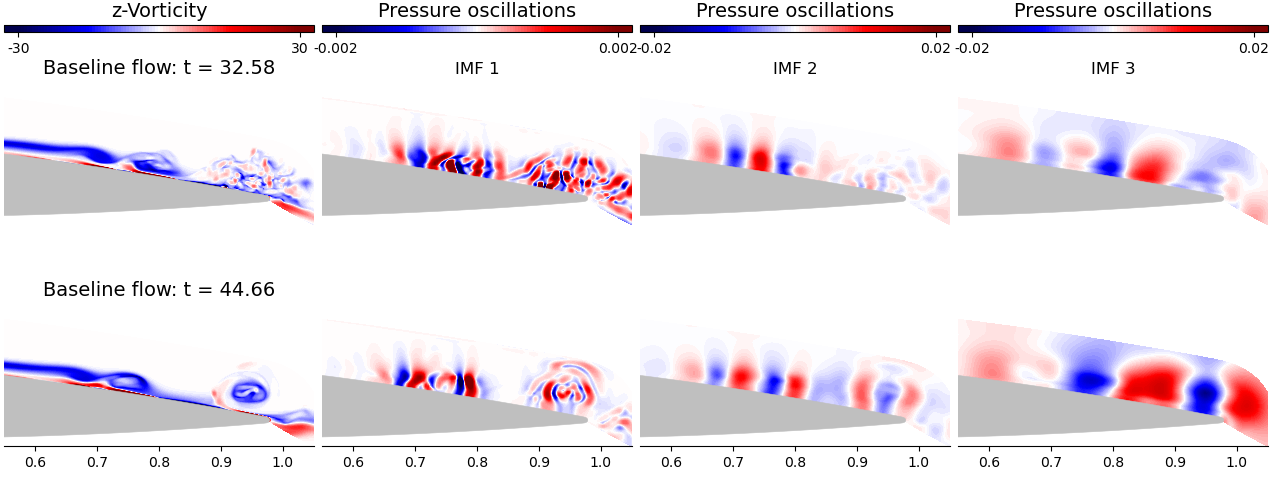}
\end{center}
\caption{EMD results of the vortex shedding mechanism on the NACA0012 suction side. The top images show the IMFs of a less correlated turbulent packet reaching the trailing edge, while the bottom ones show a coherent structure. From left to right, the contours show the z-vorticity from the LES, and the pressure oscillations from IMFs 1 to 3.}
\label{fig : Decomposition_example}
\end{figure}

\begin{figure}[b]
\begin{center}
\includegraphics[width=.99\textwidth]{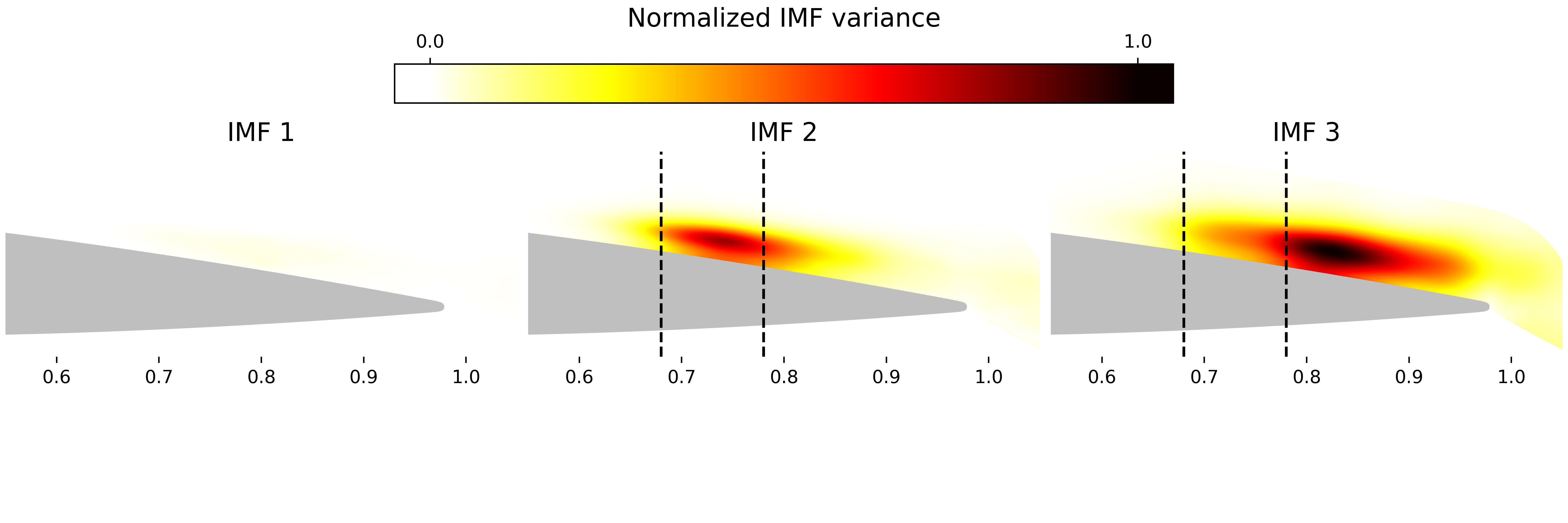}
\end{center}
\caption{Normalized variance contours computed for the time series in the IMFs at different spatial locations. From left to right, the images depict contours representing the variance data obtained from the IMFs 1 to 3. In the vortex pairing region ($0.7<x<0.8$), larger variances are observed for the IMF 2, while in the vortex advection region ($x>0.8$), IMF 3 depicts the largest variance.}
\label{fig : Imfs variance}
\end{figure}

After applying the FA-MVEMD to the instantaneous pressure data, 3 IMFs are extracted. This number of IMFs was considered adequate to separate the different scales of events related to the vortex pairing and transport of coherent structures. An example of the flow decomposition is given in Fig. \ref{fig : Decomposition_example}, where the top images show the IMFs of a less correlated turbulent packet reaching the trailing edge, while the bottom ones show a coherent structure. In the left column, the LES z-vorticity contours are plotted only to aid the visualization of the coherent structures. However, we highlight that the modal decomposition is employed to the pressure field. The distinction between the less correlated structures and the organized ones is evident in the leftmost images. The same conclusion can also be drawn from the analysis of IMFs 2 and 3 which show higher negative pressure fluctuations at the trailing edge for the coherent structures. On the other hand, when the flow is uncorrelated, the top images show weaker pressure fluctuations for the same IMFs. It is worth mentioning that the plot scales are kept with the same levels for IMFs 2 and 3, while IMF 1 is plotted with a higher saturation for visualization purposes. The figure shows that the IMF 2 brings a better representation of the structures during the pairing mechanism for  $0.7<x<0.8$. The IMF 3, in turn, provides a better representation of the coherent structures reaching the trailing edge for $x > 0.8$. These observations are reinforced by the variance contours computed for the IMFs shown in Fig. \ref{fig : Imfs variance}.

The contours in Fig. \ref{fig : Imfs variance} depict the variance of the IMFs calculated considering the time series at each spatial location. Similar to the previous analysis of the dynamic stall onset, this procedure enables us to determine the regions where the pressure variations captured by each IMF are of greater significance. The data are normalized by the largest variance among all 3 IMFs for purpose of comparison. In the figure, black dashed lines are plotted in the IMFs 2 and 3 delimiting the region where vortex pairing occurs. The contours show that the largest pressure variations due to vortex pairing are concentrated in the IMF 2. However, the largest variance is observed downstream of the pairing region in the IMF 3. In this region, larger coherent structures are advected after the vortex merging. In the following analyzes, the data obtained from the IMF 2 will be employed to investigate the vortex pairing mechanism, establishing a relation with the spectral analysis provided in Ref. \cite{ricciardi_wolf_taira_2022}.

\begin{figure}[b]
\begin{center}
\includegraphics[width=.99\textwidth]{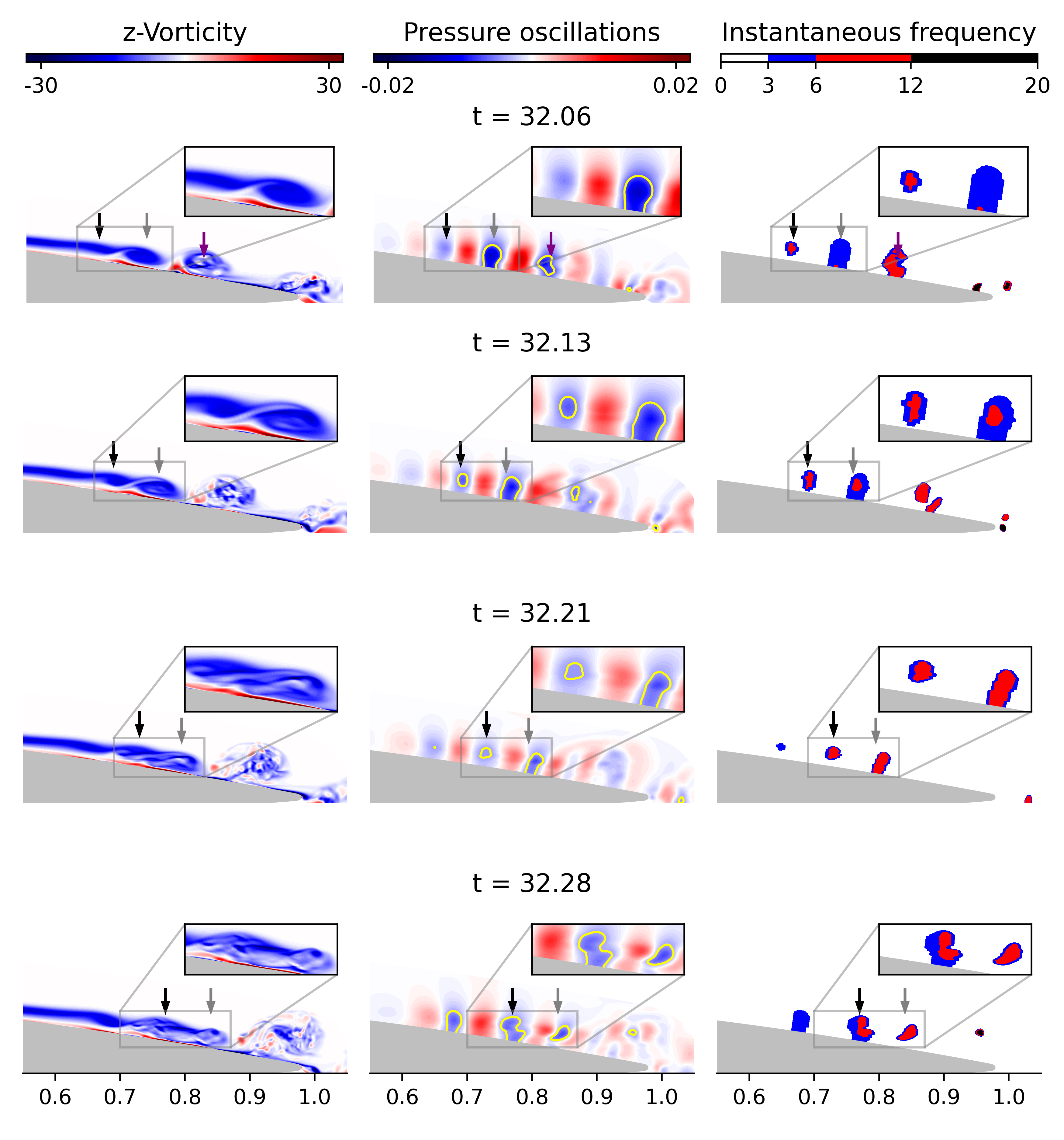}
\end{center}
\caption{Analysis of the process that originates the turbulent packet displayed at the trailing edge in Fig. \ref{fig : Decomposition_example}. From left to right the figures depict the vorticity field calculated from the baseline flow, the pressure oscillations captured by the IMF 2, and their instantaneous frequencies from the Hilbert spectral analysis.}
\label{fig : bad_pairing}
\end{figure}

\begin{figure}[htb]
\begin{center}
\includegraphics[width=.99\textwidth]{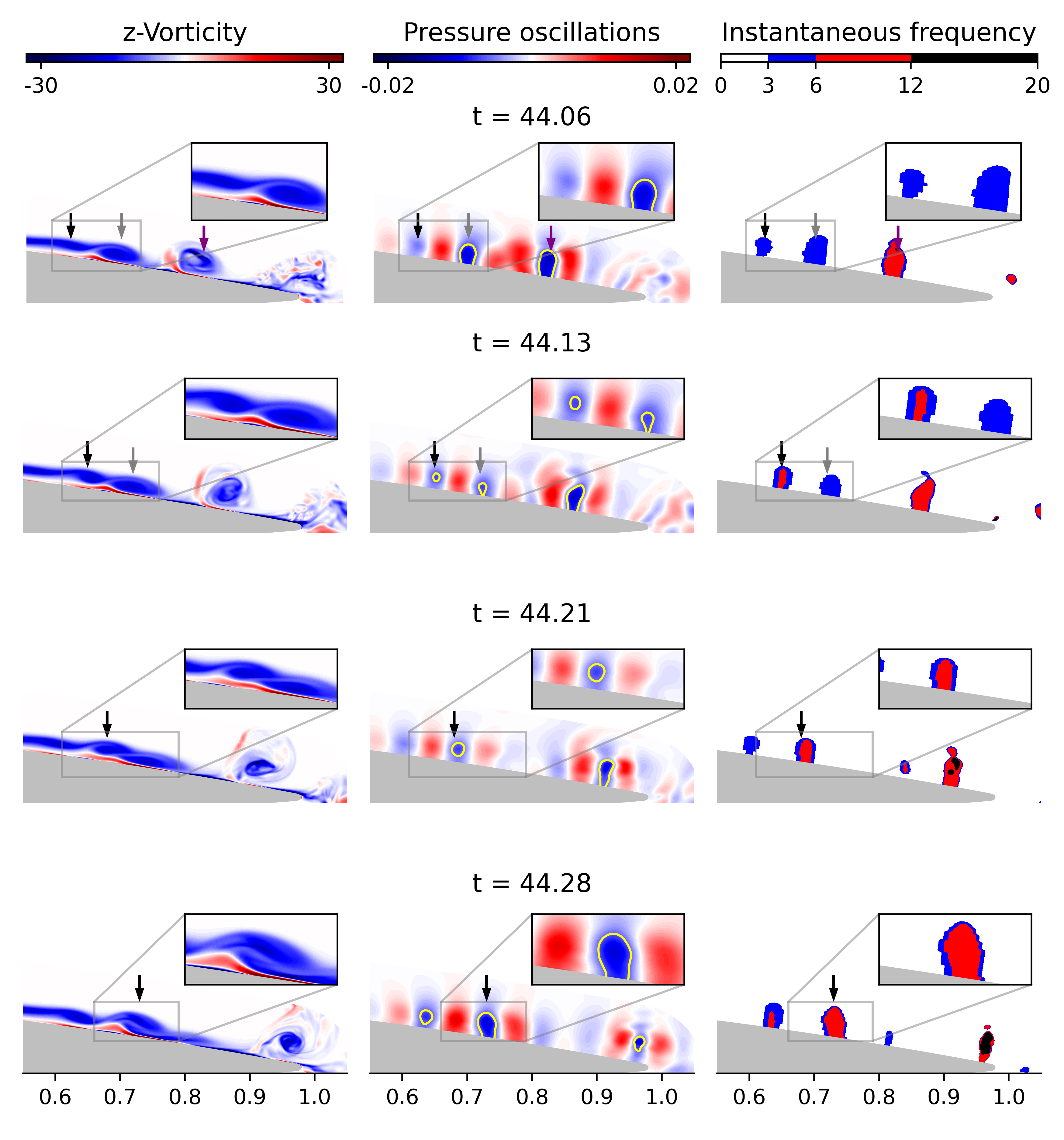}
\end{center}
\caption{Analysis of the process that originates the coherent structure displayed at the trailing edge in Fig. \ref{fig : Decomposition_example}. From left to right the figures depict the vorticity field calculated from the baseline flow, the pressure oscillations captured by the IMF 2, and their instantaneous frequencies from the Hilbert spectral analysis.}
\label{fig : good_pairing}
\end{figure}

The vortex pairing process that gives birth to the structures arriving at the trailing edge in Fig. \ref{fig : Decomposition_example} can be seen in Figs. \ref{fig : bad_pairing} and \ref{fig : good_pairing}. These figures show the cases that originate a turbulent packet and a coherent structure, respectively. According to Ref. \cite{ricciardi_wolf_taira_2022}, in the successful pairing process, constructive interference of frequencies in the range $3 \leq St \leq 6$ results in two-dimensional vortex structures downstream from the laminar separation bubble. On the other hand, when the process is unsuccessful the vortices are broken in uncorrelated turbulence. The plots show vorticity contours on the left, pressure oscillations represented by the IMF 2 in the center, and instantaneous frequencies calculated by the Hilbert spectral analysis of the same IMF in the right. In the right contours, the blue color emphasizes the regions with frequencies corresponding to the range where constructive interference is important for a successful pairing, according to Ref. \cite{ricciardi_wolf_taira_2022}. Black and gray arrows indicate the vortices that are pairing, with a third purple arrow indicating the position of the nearest downstream vortex in the beginning of the process. These arrows will be used further as a reference for comparison of the vortex positions in the IMFs and in the mrDMD modes. The temporal evolution is given by the sequence of images from top to bottom. Moreover, in the contours of the IMF 2, yellow lines delimit the regions where the pressure values are within the peak 50$\%$, being used as a reference of the vortex core size as it is considered to be a key parameter in the vortex merging process.


The unsuccessful pairing process of Fig. \ref{fig : bad_pairing} generates a less correlated turbulent packet that reaches the trailing edge as shown in Fig. \ref{fig : Decomposition_example}. The process starts with the center of the paired vortices at approximately $x = 0.66$ and 0.74, with a third vortex close to these two, at $x = 0.83$, as indicated by the arrows at time $t = 32.06$. At this instant, the instantaneous frequency of the vortex indicated by the gray arrow is in the range $3 < St < 6$, as shown by the Hilbert spectral analysis, while the leftmost vortex, indicated by the black arrow, presents a higher instantaneous frequency. In the pressure oscillations delimited by the figure inset, in turn, it can be noticed that the pressure fluctuation of the vortex pointed by the gray arrow is higher than that of the left vortex, indicated by the black arrow. At the next instant, at $t = 32.13$, the pressure oscillation contours indicate that the vortex indicated the black arrow experiences a suction increase, but the right vortex still has a relatively larger suction core. From the instantaneous frequency contours, we verify that the  vortex indicated by the gray arrow presents a considerable increase in the instantaneous frequency, as evidenced by the red color that marks regions where $6 < St < 12$ in its core. This frequency increase is associated with the advection of the vortex, which experiences a streamwise acceleration. In the following instants, at $t = 32.21$ and $32.28$, it is noticed that the pressure signature of the vortices become more diffuse and the vorticity contours show less organized structures. In this figure, the downstream vortex (indicated by the purple arrow in the first row) is situated in close proximity to the pairing vortices, resulting in the advection of all three structures as a collective entity.
%
%

The contours in Fig. \ref{fig : good_pairing} describe the birth of the coherent structure that reaches the trailing edge, as depicted in Fig. \ref{fig : Decomposition_example}. At $t = 44.06$, the pairing process starts with a stronger vortex identified by the gray arrow at $x = 0.7$ and a weaker vortex being formed upstream, at $x = 0.62$. This can be seen in the figure inset which highlights the pressure contours of the IMF 2. A third vortex marked by a purple arrow can also be seen further downstream at $x = 0.83$. At this moment, the Hilbert spectral analysis shows that both upstream vortices have instantaneous frequencies in the range $3<St<6$. At $t = 44.13$, the gray arrow marks a weakened right vortex, represented by the flattening of the vorticity contour as well as a reduction of its low pressure core, shown in the IMF 2. This is counterbalanced by a strengthening of the left vortex marked by the black arrow. At this instant, the instantaneous frequency contours show that this process occurs in a quasi-stationary manner for the right vortex (marked by the gray line), which keeps its frequency content, while the left vortex experiences a frequency increase that can be explained by a streamwise acceleration. In this case, the convective stage of the merging process occurs successfully. Here, this is noticed by the instantaneous frequency contours between $t = 44.06$ and $t = 44.13$, in which the 
vortex indicated by the gray arrow appears quasi-stationary, while the leftmost vortex accelerates, catching up with the slower one downstream. 
While the pairing occurs, the downstream vortex (initially marked by the purple arrow) is transported through the airfoil surface, distancing from the merging structure.

The dynamics portrayed in Fig. \ref{fig : good_pairing} exhibits additional similarities to the canonical cases of vortex merging described in the literature. For instance, the interaction between a vortex pair and a surface leads to an adverse pressure gradient that induces separation of the boundary layer, originating a vorticity layer of opposite sign \cite{Hirsch_1978, crow_1977, Frank_2007}. This layer promotes the pairing of co-rotating vortices by pushing the near-wall eddy upwards \cite{WANG2016116}. In the present flow, a small region of positive vorticity can be observed in the region between the pairing eddies. This region becomes larger during the lift-up of the left vortex during its motion to overcome the right eddy, as can be seen at the instants $t = 44.21$ and $t = 44.28$. In addition, the increase of the vortex core strength, displayed by the yellow line in the IMF 2 from $t = 44.21$ to 44.28, can be interpreted as the final stage of growth by diffusion as described by Ref. 
\cite{cerretelli_williamson_2003}. Furthermore, the disparity in the size of the vortex cores during the unsuccessful process, evidenced by the contours of IMF 2 at $t = 32.13$ and $32.21$ in Fig. \ref{fig : bad_pairing}, may indicate an inelastic interaction that rarely produces a single vortex, while at $t = 44.13$ the vortices appear to have similar sizes. In this latter case, vortex merging is more likely to occur according to Ref. \cite{Dritschel_1992}.
\begin{figure}[!h]
\begin{center}
\includegraphics[width=.99\textwidth]{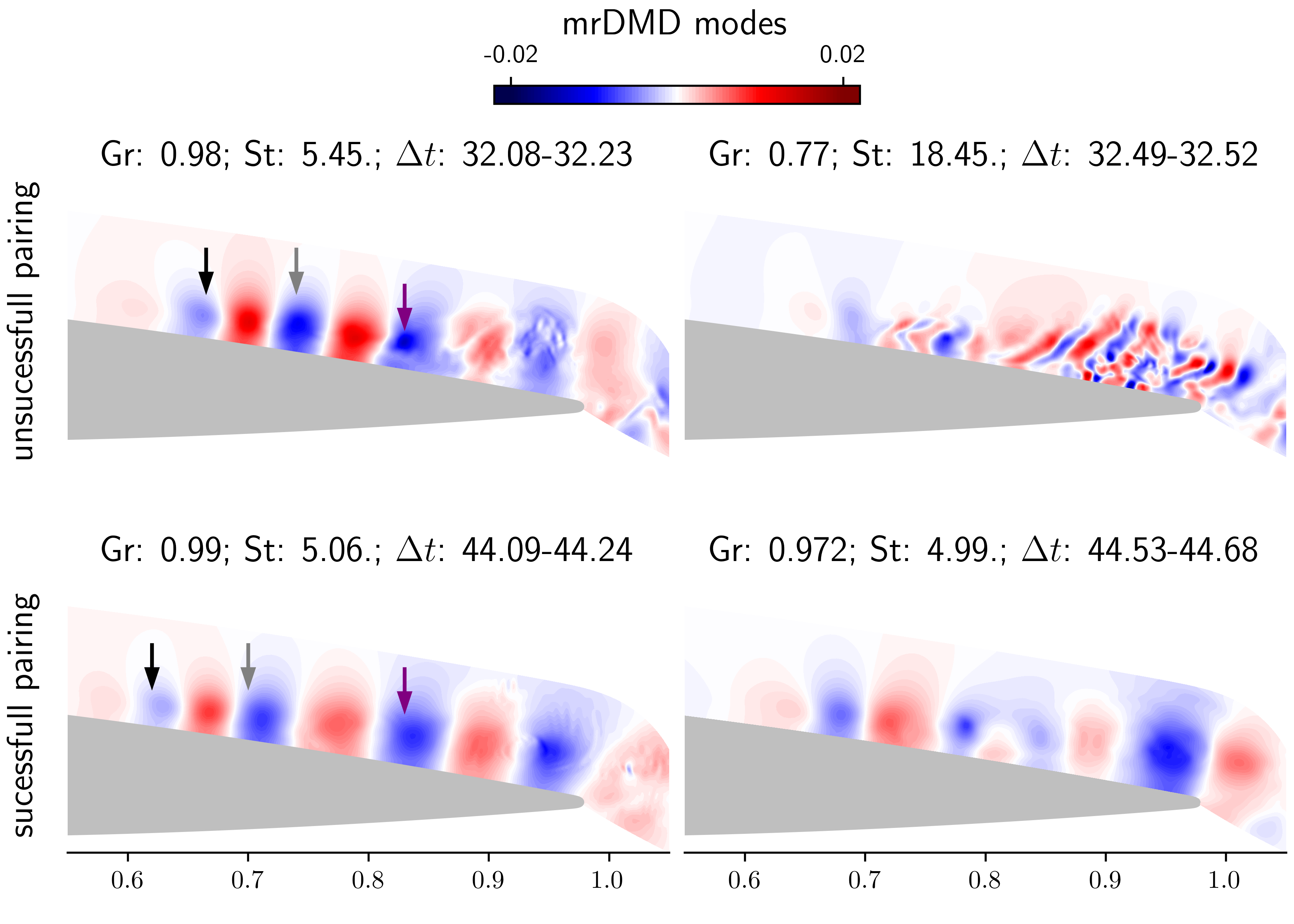}
\end{center}
\caption{Contours of the mrDMD modes representing the  pairing processes (left column) and the structures resulting from them (right column). The top row depicts the /blue{unsuccessful} pairing, while the bottom one shows the successful case. The arrows indicate the initial positions of the vortices at the beginning of each pairing process.}
\label{fig : mrDMD}
\end{figure}

The mrDMD modes that correspond to the events previously described are shown in Fig. \ref{fig : mrDMD}. In this figure, the top contours show the modes related to the unsuccessful pairing process that culminates in the arrival of a turbulent packet at the trailing edge. On the other hand, the bottom contours show modes that correspond to the instants in which the pairing process is successful, resulting in a coherent structure. The first column of the figure shows the contours for the time intervals represented in Figs. \ref{fig : bad_pairing} and \ref{fig : good_pairing}. Here, arrows are plotted in the same positions of those from these previous figures at $t = 32.06$ and $44.06$, marking the initial positions of the two vortices being paired (black and gray arrows) and the downstream vortex (purple arrow). The second column includes the instants from Fig. \ref{fig : Decomposition_example}, when the resulting coherent structures or turbulent packets reach the trailing edge. 

It is possible to see in the left column that the mrDMD modes capture the initial vortex positions at frequencies similar to those computed from the Hilbert spectral analysis at $t = 32.06$ and $44.06$. One should note that the small shift in the position of the second vortex in the successful pairing is due to the fact that it represents the initial condition at $t = 44.09$ in the mrDMD, while Fig. \ref{fig : good_pairing} shows the instant at $t = 44.06$ for the EMD. On the right column, the mode that corresponds to the coherent structure shown in Fig. \ref{fig : Decomposition_example} presents $St \approx 5$, while the turbulent packet is better represented by higher frequencies. Moreover, the DMD mode shown in the top right corner is very similar to the IMF 1 shown in Fig. \ref{fig : Decomposition_example} for $t = 32.58$. These results are also in line with what was observed in Ref. \cite{ricciardi_wolf_taira_2022} where, by means of wavelet transforms computed for a probe near the trailing edge, it was found that the transport of coherent vortices is associated with frequencies in the range $3< St < 6$. Nevertheless, with the mrDMD decomposition the depiction of the pairing phenomenon remains unclear due to the difficulty to track structures in real-time, as seen in the EMD. The fixed frequency of the mrDMD modes hinders the observation of frequency variations crucial for portraying the convective stage of the vortex pairing. 
%
%
Moreover, since the mrDMD modes do not exhibit discernible differences in the vortex pairing region, analyzing the individual modes does not provide sufficient grounds to draw further conclusions regarding the vortex pairing process.

\section{Conclusions}\label{sec3}

Modal analyses of intermittent and transient flows are carried out via the FA-MVEMD and mrDMD. Studies are conducted 
for a periodic plunging SD7003 airfoil under deep dynamic stall conditions, as well as a stationary NACA0012 at a transitional Reynolds number. Both methods of flow modal decomposition are compared on their ability to extract physically meaningful features of the different stages of the dynamic stall event including the onset of the DSV, its advection, and the trailing edge vortex formation, alongside the intermittent vortex pairing mechanism which occurs for the stationary airfoil. 

Results show that both methods successfully represent the multiple coherent structures 
and their associated frequencies related to the 
different stages of the investigated flows. However, the EMD has an advantage over the mrDMD. The former is capable of condensing a larger volume of information within a single IMF, thereby providing a more direct and convenient way of analyzing the unsteady flowfield data. Moreover, the limitation of SVD-based methods to provide low-rank representations of traveling flow structures 
is observed even when using the multi-resolution variant of the DMD. Since the lower levels of the mrDMD approach are responsible to capture the large-scale low-frequency flow structures, a large time window is employed for the decomposition. Consequently, the modes that depict
the low-frequency structures are subject to the same limitations as the standard DMD algorithm. In other words, the method assumes that such coherent structures exist throughout the entire temporal window, producing modes that 
represent sequential events in an unphysical simultaneous fashion. 

On the other hand, in the multidimensional EMD, the coherent structures depicted in the IMFs are represented with a more compact spatial support, with a few localized spurious pressure oscillations around the vortices. Such oscillations are attributed to the mathematical artifact of the method to produce signals with a well behaved Hilbert transform. In this context, it is also worth mentioning that the combination of EMD with the Hilbert spectral analysis also allows the characterization of instantaneous frequencies of the coherent structures. Therefore, this approach allows the analysis of sequential events that may occur in intermittent and transient flows. It should be highlighted that the present comparisons are focused mainly in the ability of the methods to provide physically interpretable modes. In terms of data generated by the methods, each IMF and the residue produced by the multidimensional EMD have the same size as the original data used in the decomposition, increasing the total amount of data. For the mrDMD decomposition, in turn, the size of an individual mode is that of the spatial domain.

Results of the periodic plunging airfoil show that both methods are able to identify coherent structures on the leading edge that have frequencies within the range in which actuation would have the effect of disrupting the DSV according to \citet{ramos2019active}. In this sense, the methods are able to identify the coherent structures responsible for the onset of the DSV, including the process of vortex coalescence in the leading edge. The spatial support of the traveling structures during the early stages of the DSV development is more accurately captured by the EMD in comparison to the mrDMD. At later stages, an IMF highlights the feeding sheet that connects the DSV to the leading edge and the small vortical structures within the DSV, suggesting that the coalescence process that gives birth to the DSV persists through later stages of the dynamic stall process. Also in the multidimensional EMD, both the trailing edge vortex and the DSV are well captured by an IMF and the residue in the instant when the former ejects the latter from the airfoil surface. Multiple DMD modes would be required to represent the same features at different frequencies.

Finally, in the stationary NACA0012 case, the EMD provides a better characterization of intermittent features such as the vortex pairing taking place in the airfoil suction side. An analysis performed with this method allows for the identification of similarities between the successful vortex pairing on the present case and that of co-rotating vortices in ground proximity, as described by Refs. \cite{crow_1977, Frank_2007, WANG2016116}. Through the combination of the multidimensional EMD and Hilbert spectral analysis, the impact of the vortex intensities and their instantaneous frequencies is shown in the pairing process. The mrDMD modes, in turn, are limited to information about the initial condition and a fixed frequency, failing to capture the frequency variations that characterize the convective stage of vortex merging. Even so, the notorious similarity between the isolated structures and frequencies obtained by both methods and the analysis carried out by  \citet{ricciardi_wolf_taira_2022} corroborate their validation.



\section*{Declarations}
\subsection*{Ethical Approval}
Not applicable.
\subsection*{Competing interests}
The authors declare that they have no competing interests.
\subsection*{Authors' contributions}
\begin{itemize}
    \item L.S. conceptualization, methodology, software development, validation, formal analysis, investigation, writing original draft;
    \item R.M. conceptualization, methodology, software development, formal analysis, supervision, writing \& review;
    \item W.W. conceptualization, methodology, formal analysis, supervision, resources, writing \& review;
\end{itemize}
\subsection*{Funding}
Fundação de Amparo à Pesquisa do Estado de São Paulo (FAPESP): Grants No. 2013/08293-7, 2013/07375-0, 2021/06448-0, 2022/08567-9 and 2022/09196-4.
\subsection*{Availability of data and materials}
Datasets are available as requested. Numerical tools are available from the external repository links available in the manuscript.
\subsection*{Acknowledgments}
CEPID-CeMEAI and CENAPAD-SP are acknowledged for providing the computational resources for this research through the Euler and Lovelace clusters, respectively.





\bibliography{sn-bibliography}

\end{document}